\documentclass[a4paper,11pt]{article}

\usepackage{color}
\usepackage{ulem}
\usepackage{hyperref}
\usepackage{enumerate}
\usepackage{amsmath}
\usepackage{graphicx}

\usepackage{float}
\usepackage{amssymb}
\usepackage{subfig}
\makeatletter

\newcommand{\Rmnum}[1]{\expandafter\@slowromancap\romannumeral #1@}
\makeatother

\def \be {\begin{equation}}
\def \ee {\end{equation}}
\def \ba {\begin{array}}
\def \ea {\end{array}}
\def \bea {\begin{eqnarray}}
\def \eea {\end{eqnarray}}

\def \ble {\begin{widetext}\begin{equation}}
\def \ele {\end{equation}\end{widetext}}
\def \blea {\begin{widetext}\begin{eqnarray}}
\def \elea {\end{eqnarray}\end{widetext}}

\def \nn {\nonumber}

\def \ketphi {|\phi\rangle}

\def \transA {\mathcal{T}_A^{\psi|\phi}}

\def \blea {\begin{widetext}\begin{eqnarray}}
\def \elea {\end{eqnarray}\end{widetext}}
\def \mO {\mathcal{O}}

\def \p {\partial}
\begin{document}

\title{Pseudo entropy and pseudo-Hermiticity in quantum field theories}

%\author{Wu-zhong Guo}
%\email{wuzhong@hust.edu.cn}
%\affiliation{School of Physics, Huazhong University of Science and Technology,
%Luoyu Road 1037, Wuhan, Hubei 430074, China}

\author{Wu-zhong Guo\footnote{wuzhong@hust.edu.cn},   Yaozong Jiang\footnote{d202280080@hust.edu.cn}}

\date{}
\maketitle

\vspace{-10mm}
\begin{center}
{\it School of Physics, Huazhong University of Science and Technology,\\
 Wuhan, Hubei
430074, China
\vspace{1mm}
}
\vspace{10mm}
\end{center}

\begin{abstract}
%In this paper we investigate pseudo R\'enyi entropy in quantum field theoires.  The transition matrix is constructed by acting operators located in different regions on the vacuum  . If the two operators are located in the left and right Rindler wedges respectively, it is found the logarithmic term of the pseudo R\'enyi entropy should be real. Otherwise the result may be complex. We directly evaluate some examples in 2-dimensional conformal field theories (CFTs). Further, we find it can be explained by pseudo-Hermitian condition. Moreover, we also study the divergent term of pseudo R\'enyi entropy. It is found the second pseudo R\'enyi entropy has a universal divergent term  in 2-dimensional CFTs, which is only associated with the conformal dimension of the operator. For $n$-th ($n\ge 3$) pseudo R\'enyi entropy the divergent term is related to the details of the theory.

In this paper, we explore the concept of pseudo Rényi entropy within the context of quantum field theories (QFTs). The transition matrix is constructed by applying operators situated in different regions to the vacuum state. Specifically, when the operators are positioned in the left and right Rindler wedges respectively, we discover that the logarithmic term of the pseudo Rényi entropy is necessarily real. In other cases, the result might be complex. We provide direct evaluations of specific examples within 2-dimensional conformal field theories (CFTs). Furthermore, we establish a connection between these findings and the pseudo-Hermitian condition. Our analysis reveals that the reality or complexity of the logarithmic term of pseudo R\'enyi entropy can be explained through this pseudo-Hermitian framework.

Additionally, we investigate the divergent term of the pseudo Rényi entropy. Interestingly, we observe a universal divergent term in the second pseudo Rényi entropy within 2-dimensional CFTs. This universal term is solely dependent on the conformal dimension of the operator under consideration. For $n$-th pseudo Rényi entropy ($n\ge 3$), the divergent term is intricately related to the specific details of the underlying theory.
\end{abstract}

\maketitle

\section{Introduction}
Density matrix is a fundamental concept in quantum mechanics, used to describe the states of a given system. 
%Generally, the density matrix can be written as $\rho=\sum p_i|\phi_i\rangle \langle \phi_i|$, where $|\phi_i\rangle$ are pure states, $p_i$ are positive coefficients.  
The reduced density matrix plays a crucial role in characterizing quantum correlations or entanglement between subsystems of the given system. One could define various relevant quantities, employed as a function of the reduced density matrix, serves as a measure of entanglement.
%The von Neumann entropy, employed as a function of the density matrix, serves as a measure to quantify the entanglement within a system. 

In quantum field theories (QFTs), some of the entanglement measure can be well-defined and computed analytically or numerically. In the context of AdS/CFT\cite{Maldacena:1997re}-\cite{Witten:1998qj} it is also found certain quantities have some nice gravity dual such as entanglement entropy (EE)\cite{Ryu:2006bv}\cite{Hubeny:2007xt}, Entanglement of purification (EoP)\cite{Takayanagi:2017knl}\cite{Nguyen:2017yqw}, negativity\cite{Kudler-Flam:2018qjo}\cite{Kusuki:2019zsp}, reflected entropy\cite{Dutta:2019gen}, R\'enyi entropy\cite{Dong:2016fnf} etc. These studies enable us to gain further insights into the relation between entanglement and geometry\cite{VanRaamsdonk:2010pw}\cite{Almheiri:2014lwa}.

The quantities mentioned above can all be regarded as functions of the reduced density matrix.  One could also generalize the density matrix to the transition matrix, which involves two different states $|\phi\rangle$ and $|\psi\rangle$. Without normalization it can be taken as the operator $|\psi\rangle \langle \phi|$. Actually, in many cases we have already used the transition matrix. For example,  the expectation value of an operator $O$ in the state $|\psi\rangle$ is $\langle \psi| O|\psi\rangle =tr (|\psi\rangle \langle \psi| O)$ can be taken as the trace of the operator $|\psi\rangle \langle \phi|$ with $|\phi\rangle=O^\dagger |\psi\rangle$. 

Similar to the density matrix, the concept of reduced transfer matrices can be introduced by replacing the trace operation with partial traces for a given subsystem $A$ in certain contexts. That is
\bea
\transA:= tr_{\bar A} \frac{|\psi\rangle \langle \phi|}{\langle  \phi|\psi\rangle},
\eea
where $\bar A$ is the complementary part of $A$. In \cite{Nakata:2020luh} the authors introduce the so-called pseudo entropy as a new generalization of EE, which is the von Neumann entropy of the operator $\transA$,
\bea
S(\transA)= - tr_A ( \transA \log \transA ).
\eea
See also the similar quantity defined in \cite{Murciano:2021dga}. It is interesting that pseudo entropy also has a gravity dual similar as EE if the transition matrix $|\psi\rangle \langle \phi|$ has a bulk geometry dual.  To evaluate pseudo entropy we usually calculate its one parameter generalization  pseudo R\'enyi entropy, defined as
\bea
S^{(n)}(\transA)= \frac{\log tr_A (\transA)^n}{1-n},
\eea
where $n$ is an integer. 

The reduced transition matrix is generally non-hermitian, thus the eigenvalues of it may be complex. The pseudo R\'enyi entropy and pseudo entropy may also be complex number. The imaginary part of pseudo entropy can be explained as timelike entanglement\cite{Doi:2022iyj}.  It is interesting that the pseudo R\'enyi entropy can be connected with the R\'enyi entropy for the superposition states of $|\psi\rangle$ and $|\phi\rangle$ by a sum rule\cite{Guo:2023aio}. 
There have been many recent studies related to pseudo-entropy, please refer to \cite{Guo:2022jzs}-\cite{Shinmyo:2023eci}. 

The class of the transition matrix that has real-valued  pseudo entropy should only be a special subset of the transition matrix. Motivated by the recent works on nonhermitian physics \cite{Ashida:2020dkc, Bender:2007nj, Mostafazadeh:2008pw}, the authors in \cite{Guo:2022jzs} find the real-valued condition of pseudo entropy can be understood by the concept of pseudo hermitian. This paper represents further research on the topices discussed above. In QFTs the states are constructed by acting operators on the vacuum states. The transition matrix usually includes two different operators, thus the pseudo R\'enyi entropy should include more information on the correlators than the EE. Therefore, we expect the real-valued condition of pseudo R\'enyi entropy can be related to some properties of correlation functions in QFTs.

In this paper we mainly focus on the excited states constructed by acting local operators in Minkowski spacetime. Some examples are discussed in the paper\cite{Guo:2022jzs}. We extend the results to more general cases and find some new properties of the pseudo R\'enyi entropy. We also explain our new results by pseudo hermitian condition.  Besides, the pseudo entropy  is usually divergent if the operators are located near the lightcone\cite{Guo:2022sfl}. We show that the divergence is universal for the second pseudo R\'enyi entropy, which only depends on the conformal dimension of the operator for 2-dimensional conformal field theories. But for $n\ge 3$ the divergent terms depend on more details of the theory.

The paper is organized as follows.  Section.\ref{sectionset} is the general set-up, which includes the transition matrix that we will consider. In section.\ref{sectionresult} we will firstly evaluate the pseudo R\'enyi entropy for some examples for different cases.  Then we will analyse the result for general operator by using properties of correlation functions. In section.\ref{sectionlightcone} we focus on the divergent term for operators located near the lightcone. For pseudo R\'enyi entropy we find the divergent term is universal, which only depends on the conformal dimension of the operator near the lightcone. Section.\ref{sectionpseudo} is devoted to pseudo hermitian condition for the examples that we discussed in previous section. The results can be explained by the pseudo hermitian condition. The last section is the conclusion. In the appendices we show more details of the calculations.

\section{Transition matrix construction in QFTs}\label{sectionset}
For a given subsystem $A$, the local operator algebra  $\mathcal{R}(A)$ consists of the operators supported in $\mathcal{A}$, where $\mathcal{D}(\mathcal{A})$ is the domain of dependence of $A$. Let $\tilde{A}$ be another subsystem which is spacelike with $A$. Denote the algebra associated with $\tilde{A}$ by $\mathcal{R}(\tilde{A})$. By microcausility  we would have 
$[\mO,\tilde{\mO}]=0$ for $\mO\in \mathcal{R}(A)$ and $\tilde{\mO} \in \mathcal{R}(\tilde{A})$. 
According to the Reeh-Schlieder theorem\cite{Hagg}\cite{Witten:2018zxz}, for any pure state $\ketphi$ there exists local operators $\mathcal{O}_\phi$  in $\mathcal{R}(A)$ such that $\ketphi$  can be approximated by $\mathcal{O}_\phi| 0\rangle$, that is the distance between $\mathcal{O}_\phi| 0\rangle$ and $\ketphi$ can be arbitrary small. Therefore, we would like to consider the general transition matrix defined as 
\bea
\mathcal{T}^{\mO| \mO'}:= \frac{\mO|0\rangle \langle 0| \mO' }{\langle \mO \mO' \rangle},
\eea
where $\mO \in \mathcal{R}(A)$ and $\mO'\in \mathcal{R}(A')$ for subsystem $A$ and $A'$. We would assume $\mO,\mO'$ are Hermitian operators. Any operators can be written as linear combination of  two Hermitian operators. If the operators $\mO,\mO'$ are not Hermitian, one could rewrite the transition matrix as linear combinations of transition matrices constructed by hermitian operators.

In this paper we would mainly focus on the Rindler wedges in Minkowski spacetime. For d-dimensional spacetime the metric is  $ds^2=-dt^2+dx^2+d\vec{y}^2$, where $\vec{x}'$ are coordinates of $d$-dimensional Euclidean space. The left Rindler wedge is defined in the region $|t|<-x$. The right Rindler wedge satisfies the condition $|t|<x$. The region $t>|x|$ ($t<-|x|$) is called the expanding (contracting) degenerate Kasner universe. The Minkowski vacuum $|0\rangle$ can be written as entangled states between the left and right Rindler wedges\cite{Unruh:1976db}.  If taking $x\in(0,+\infty)$ as the subsystem $A$, one could calculate the reduced density matrix $\rho^0_A:= tr_{\bar A}|0\rangle \langle 0|$, where $\bar A$ is the region $x\in (-\infty,0)$, and define the R\'enyi entropy of $A$, which are divergent due to infinite size of the subsystem. But  one could focus on the difference  
\bea
\Delta S^{(n)}=S^{(n)}(\mathcal{T}_A^{\mO| \mO'})-S^{(n)}(\rho^0_A),
\eea
which is usually finite. 

Let us consider the local operators $\mO=O(t,\vec{x})$ and $\mO'=O(t',\vec{x}')$.
The transition matrix is given by
\bea\label{transitionlocal}
\mathcal{T}^{O| O'}:= \frac{O(t,\vec{x})|0\rangle \langle 0| O(t',\vec{x}') }{\langle O(t',\vec{x}') O(t,\vec{x})\rangle},
\eea
We assume the operator $O$ is hermitian.
One could use replica method to evaluate pseudo R\'enyi entropy. We will consider the operators  $O$ and $O'$ are located in different regions, e.g., $O$ is in the left Rindler wedge and $O'$ is in the right Rindler wedge. The positions of the operators play a crucial role in determining the behavior of pseudo R\'enyi entropy. It is obvious the pseudo R\'enyi entropy can be used as a tool to detect the spectra of $\transA$.
In this paper we will mainly focus on primary operator $O$ in 2-dimensional CFTs. 
In the following we will firstly calculate the pseudo R\'enyi entropy for some special cases.
Let us fix the coordinate $(t',x')=(0,x')$ with $x'>0$, i.e.,  $O(t',x')$  is in the right Rindler wedge and on the time slice $t=0$. 
We will mainly focus on the following cases:\\

\text{Case I:}  $x<0,t=0$,  i.e.,  $O(t,x)$ is on the time slice $t=0$. \\

\text{Case II:} $x<t<-x $ and $t\ne 0$,  i.e.,  $O(t,x)$ is in the left Rindler wedge.\\

\text{Case III:}  $-t<x<t$,  i.e., $O(t,x)$ is in the expanding degenerate Kasner universe.\\

\begin{figure}[H]
\centering
\includegraphics[scale=0.4]{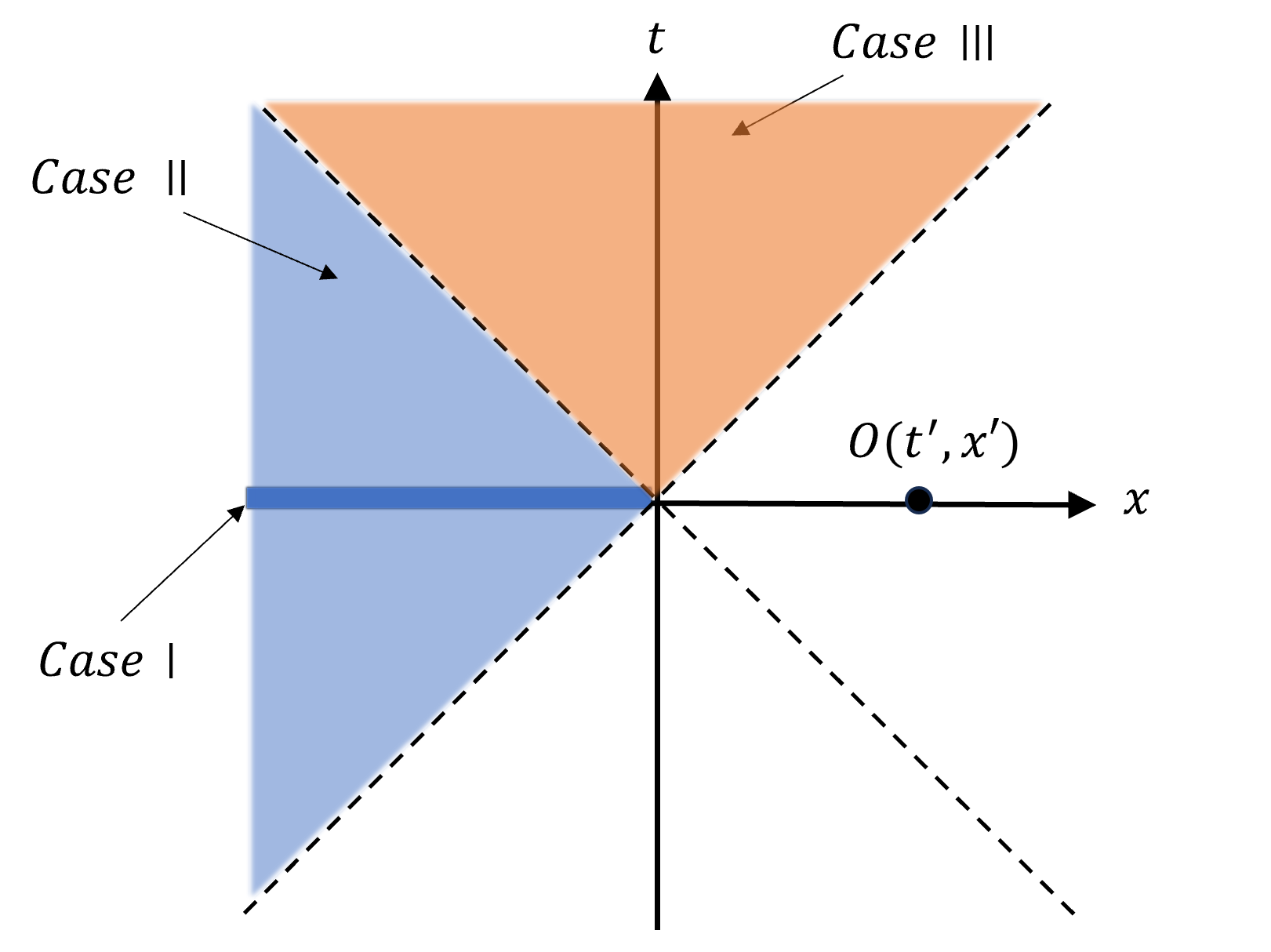}
\caption{Three different cases that we consider in this paper. The operator $O(t',x')$ is fixed at the point $(0,x')$ with $x'>0$. $O(t,x)$ located in the blue region (case I), light blue region (case II) and orange (case III).}
\label{fig:threecases}
\end{figure}

In Figure.\ref{fig:threecases} we show the three different cases. 

%~\\
%\textit{Case I}: $O(t,\vec{x})$ and $O(t',\vec{x}')$ are both located in the left Rindler wedge. \\
%~\\
%\textit{Case II}: $O(t,\vec{x})$ is located in the left Rindler wedge.   $O(t',\vec{x}')$ is located in right Rindler wedge. \\
%~\\
%\textit{Case III}: $O(t,\vec{x})$ and $O(t',\vec{x}')$ are located in the Kasner universe.\\

%In the following we would like to prove that the transition matrices in some cases have real eigenvalues, or complex ones coming in complex  conjugate pairs. According to \cite{} it is equal to the condition that the transition matrices are $\eta$-pseudo Hermitian with $\eta=\eta_A \otimes \eta_{\bar A}$. \\

\section{Pseudo R\'enyi entropy }\label{sectionresult}
In this section we will directly evaluate the pseudo R\'enyi entropy by replica method for the three different cases. 
\subsection{Review of replica method}
We will consider the transition matrix 
\bea
\mathcal{T}_E=\mathcal{N} e^{-\tau  H}O(0,x)|0\rangle \langle 0| O(0,x')e^{-\tau' H},
\eea
where $\mathcal{N}$ is the normalization constant. $\tau$ and $\tau'$ is the Euclidean time, later we will obtain the real time result by analytical continuation of $\tau$. Define the coordinates $w=x+i\tau$ and $\bar w=x-i\tau$. The transition  matrix is given by 
\bea
\mathcal{T}_E=\frac{O(w_1,\bar w_1)|0\rangle \langle 0| O(w_2,\bar w_2)}{\langle O(w_1,\bar w_1)O(w_2,\bar w_2)\rangle},
\eea
with $w_1=x-i\tau$, $\bar w_1=x+i\tau$ and $w_2=x'+i\tau'$, $\bar w_2=x'-i\tau'$. Define the reduced transition matrix $\mathcal{T}_{E,A}:=tr_{\bar A}\mathcal{T}_E $, which can be prepared by Euclidean path integral with operators inserted.  $tr (\mathcal{T}_{E,A})^n$ is given by correlators on the $n$-sheet manifold $\Sigma_n$,
\bea\label{trTncorrelator}
\frac{tr (\mathcal{T}_{E,A})^n}{tr(\rho_A^0)^n}=\frac{\langle O(w_1,\bar w_1)O(w_2,\bar w_2)...O(w_{2n-1},\bar w_{2n-1})O(w_{2n},\bar w_{2n})\rangle_{\Sigma_n}}{\langle O(w_1,\bar w_1)O(w_2,\bar w_2)\rangle_{\Sigma_1}^n},
\eea
where $O(w_{2i-1},\bar w_{2i-1})$ and $O(w_{2i},\bar w_{2i})$ with $i=1,2,...,n$ are operators inserted on $i$-sheet. In general it is hard to evaluate the $2n$-point correlation functions on the manifold $\Sigma_n$. We will mainly focus on  subsystem $A=(0,+\infty)$ in 2-dimensional CFTs. One could use the transformation $z=w^{1/n}$, the $n$-sheet manifold $\Sigma_n$ is mapped to $z$-plane. The $2n$-point correlation funcation is given by
\bea\label{mapwtoz}
&&\langle O(w_1,\bar w_1)O(w_2,\bar w_2)...O(w_{2n-1},\bar w_{2n-1})O(w_{2n},\bar w_{2n})\rangle_{\Sigma_n}\nn \\
&&=\prod_{j=1}^{2n}\left( \frac{dz_j}{dw_j}\right)^h\left( \frac{d\bar z_j}{d\bar w_j}\right)^{\bar h} \langle O(z_1,\bar z_1)O(z_2,\bar z_2)...O(z_{2n-1},\bar z_{2n-1})O(z_{2n},\bar z_{2n})\rangle,\nn
\\
\eea
where $z_j=w_j^{1/n}$ and $\bar z_j =\bar w_j^{1/n}$. By definition the variation of the psuedo R\'enyi entropy from the vacuum state is given by
\bea
\Delta S^{(n)}=\frac{1}{1-n}\log \frac{tr (\mathcal{T}_{E,A})^n}{tr(\rho_A^0)^n}.
\eea

In this paper we will consider the transition matrix (\ref{transitionlocal}) with real time. One could evaluate the pseudo R\'enyi entropy by analytical continuation $\tau=\epsilon +it $ and $\tau' =\epsilon -it'$, where $\epsilon$ is the UV cut-off. In the final result we will take the limit $\epsilon \to 0$. With this one could obtain the pseudo R\'enyi entropy for the transition matrix $\mathcal{T}_A^{O|O'}$.

For $n=2$ by the conformal transformation $z=w^{1/2}$, the coordinates $w_i\in \Sigma_{2}$  are mapped to $z_1=-z_3=\sqrt{w_1}$ and $z_2=-z_4=\sqrt{w_2}$.  We can obtain that
\bea\label{S2}
\frac{tr (\mathcal{T}_A^{O|O'})^2}{tr(\rho_A^0)^2}=(\eta (1-\eta))^{2h}(\bar \eta (1-\bar \eta))^{2\bar h} G(\eta,\bar \eta),
\eea
where the cross ratio $\eta:= \frac{(z_1-z_2)(z_3-z_4)}{(z_1-z_3)(z_2-z_4)}$ and $\bar \eta:= \frac{(\bar z_1-\bar z_2)(\bar z_3-\bar z_4)}{(\bar z_1-\bar z_3)(\bar z_2-\bar z_4)}$, which is related to the coordinate $(t,x)$ and $(t',x')$, see the Appendix.\ref{crossratiosection} for details.

In the following we will start with some simple examples and show the general properties of pseudo R\'enyi entropy for the three different cases. Then we try to extent the conclusions for more general operators.

\subsection{Two-dimensional free boson}
Let us show the pseudo R\'enyi entropy for the transition matrix (\ref{transitionlocal}) with operators $\partial \phi \bar{\partial} \phi$ and $\mathcal{V}_\alpha :=e^{i\alpha \phi}+e^{-i\alpha \phi}$. \\

\subsubsection{Operator $\partial\phi\bar{\partial}\phi$ }
\par For the operator $\partial\phi\bar{\partial}\phi$, it has conformal dimension $h=\bar{h}=1$. By using (\ref{trTncorrelator}) and (\ref{mapwtoz}), we have
\bea
&&tr (\mathcal{T}_{E,A})^n/tr(\rho_A^0)^n \nn \\
&&= \frac{\prod_{j=1}^{2n}(z_j \bar z_j)^{1-n}}{n^{4n}}\frac{\langle \p \phi (z_1)\p \phi(z_2)...\p \phi(z_{2n})\rangle\langle \bar{\p} \phi (\bar z_1)\bar{\p} \phi(\bar z_2)...\bar{\p}\phi(z_{2n})\rangle}{(\langle \p \phi(w_1)\p \phi(w_2)\rangle \langle \bar{\p }\phi(\bar w_1)\bar{\p }\phi(\bar w_2))^n}.\nn
\eea
Let us consider the second  pseudo R\'enyi entropy $\Delta S^{(2)}$. 
With some calculations we have
\bea\label{S2free1}
\frac{tr (\mathcal{T}_A^{O|O'})^2}{tr(\rho_A^0)^2}=|\eta^2+(1-\eta)^2\eta^2+(1-\eta)^2|^2.
\eea

\begin{figure}[H]
\centering
\subfloat[Case \Rmnum{1}]{
		\includegraphics[scale=0.33]{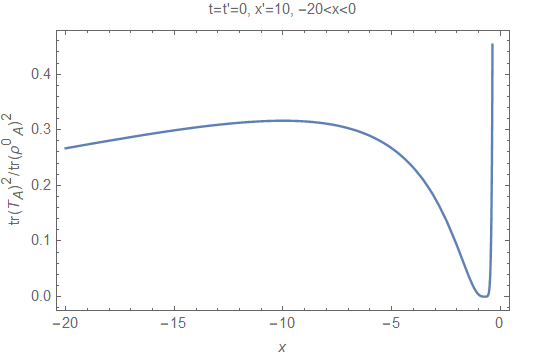}}
\subfloat[Case \Rmnum{2}]{
		\includegraphics[scale=0.33]{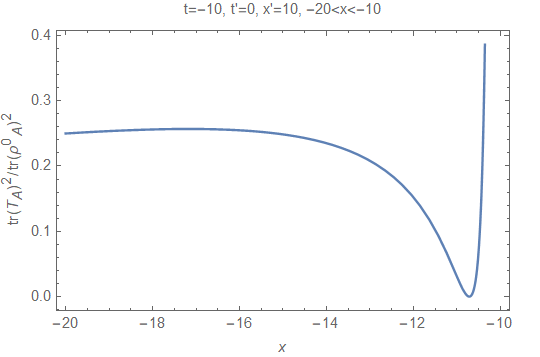}}
\\
\subfloat[Case \Rmnum{2}]{
		\includegraphics[scale=0.41]{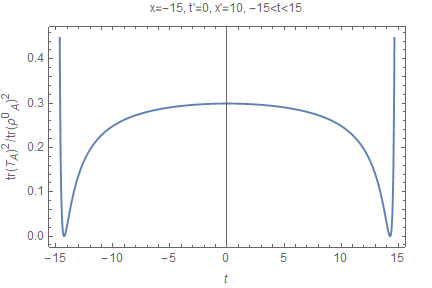}}
\subfloat[Case \Rmnum{3}]{
		\includegraphics[scale=0.41]{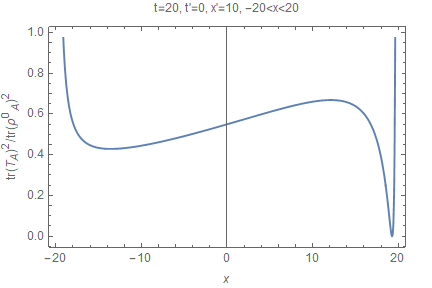}}
\caption{The plots of the logarithmic part of $\Delta S^{(2)}$ for the operator $\partial\phi\bar{\partial}\phi$.  For all the plots $(t',x')$ is fixed to be $(0,10)$. (a) is for Case I, $t=0$, $x\in (-20,0)$. (b) is for Case II, $t=-10$, $x\in(-20,-10)$. (c) is also for Case II, $x=-15$, $t\in(-15,15)$. (d) is for Case III, $t=20$, $x\in(-20,20)$.}
\label{fig:doubleparphi,n=2}
\end{figure}
As shown in Appendix.\ref{crossratiosection} the cross ration $\eta$ and $\bar \eta$ would have different values for the three different cases. Taking the cross ratios into the above formula, one could obtain the result which depends on the coordinate $(t,x)$. Some results are shown in Fig.\ref{fig:doubleparphi,n=2}. For the three cases we find the logarithmic part of $\Delta S^{(2)}$ are all positive. 

One could also calculate the $\Delta S^{(n)}$ ($n\ge 3$) by using Wick theorem for the free scalar theory. For more details of the results for free scalar theory see the Appendix.\ref{appfreescalar}, which is very similar with $\Delta S^{(2)}$.

For the operator $\partial\phi\bar{\partial}\phi$ we find the logarithmic part of $\Delta S^{(n)}$ with $n=2,3,4$ are all positive in the three cases. This implies the eigenvalues of the reduced transition matrix $\transA$ may be real or come in complex conjugate pairs. Another interesting fact is that $\Delta S^{(n)}$ is divergent when the operator $O(t,x)$ approaches to the lightcone. In the following examples we will also find the similar results.

\subsubsection{Operator $\mathcal{V}_\alpha$}\label{vertexoperatorsection}

\par For the vertex operator $\mathcal{V}_\alpha$, The conformal dimension of this operator is $h=\bar h= \frac{\alpha^2}{2}$. By using the formula for the correlation function of the vertex operator,
\begin{equation}\label{vertex,corfunction}
\langle :e^{i\alpha_1\phi(z_1,\bar{z}_1)}: \cdots: e^{i\alpha_n\phi(z_n,\bar{z}_n)}:\rangle=\prod_{i<j}^{n}{\lvert z_i-z_j\rvert}^{2\alpha_i\alpha_j},
\end{equation}

\noindent where $\sum_{i=1}^{n}\alpha_i=0$.
\par We could obtain 
\bea\label{S2freeVertex}
\frac{tr (\mathcal{T}_A^{O|O'})^2}{tr(\rho_A^0)^2}= \frac{1+(\eta \bar \eta)^{2\alpha^2}+[(1-\eta)(1-\bar \eta)]^{2\alpha^2}}{2}.
\eea
\noindent Taking the cross ratios intro the above equation, one could obtain the results, which are graphically represented in Fig. \ref{fig:vertex,n=2}.

\par We also show the result of $\Delta S^{(3)}$ in the Appendix.\ref{appfreescalar}. In Fig.\ref{fig:vertex,n=2} we show the result for three different cases with respect to the parameter $\alpha$.
\par For the operator $\mathcal{V}_\alpha$ we find that $tr_A\left(\mathcal{T}_A^{O|O'}\right)^{n}/tr_A(\rho_A^0)^{n}$ with $n=2,3$ is real for the case \Rmnum {1} and case \Rmnum {2}. But the results are generally complex in case \Rmnum {3}, which is different from the operator $\partial\phi\bar{\partial}\phi$.  We also find near the lightcone $\Delta S^{(n)}$ are also divergent.

\begin{figure}[H]
\centering
\subfloat[Case \Rmnum{1}]{
		\includegraphics[scale=0.33]{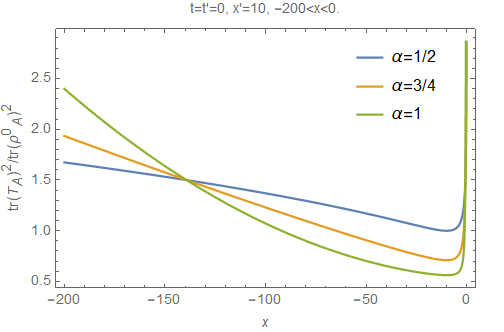}}
\subfloat[Case \Rmnum{2}]{
		\includegraphics[scale=0.33]{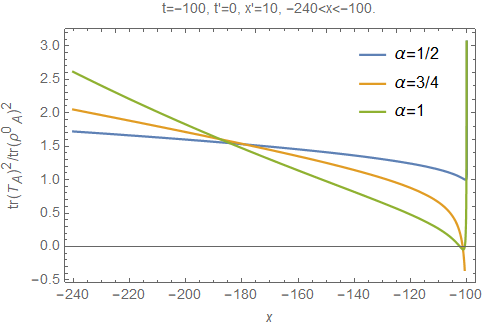}}
\\
\subfloat[Case \Rmnum{3}]{
		\includegraphics[scale=0.37]{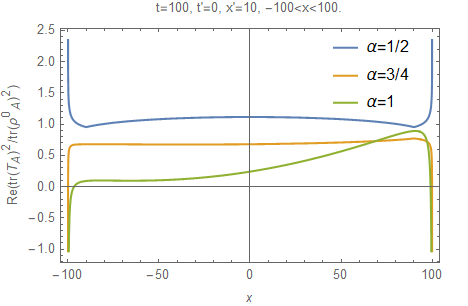}}
\subfloat[Case \Rmnum{3}]{
		\includegraphics[scale=0.37]{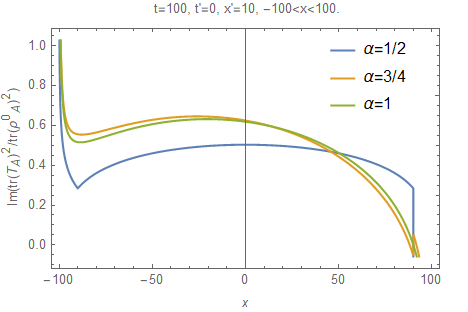}}
\caption{The plots of the logarithmic part of $\Delta S^{(2)}$ for the operator $\mathcal{V}_\alpha$ with respect to spatial position $x$ in three cases. For all the plots we fix $x'=10$. (a) is for Case I, $t=0$, $x\in(-200,0)$. (b) is for Case II, $t=-100$, $x\in(-240.-100)$. (c) and (d) are for Case III, we fix $t=100$, $x\in(-100,100)$. The results are complex, (c) and (d) are respectively the real and imaginary part of  $tr (\mathcal{T}_A^{O|O'})^2/tr(\rho_A^0)^2$. }
\label{fig:vertex,n=2}
\end{figure}

\subsection{Minimal model}\label{minimalmodelsection}
Let us consider another simple example: the operator $\phi_{(r,s)}$ in the minimal model $\mathcal{M}(p,q)$. The conformal dimension of this operator is
\begin{equation}
h=\bar{h}=\frac{(pr-qs)^2-(p-q)^2}{4pq},
\end{equation}
We would like to consider the operator $\phi_{(2,1)}$, the conformal dimension is $h=\bar h=\frac{3p}{4q}-\frac{1}{2}$. To evaluate $tr (\mathcal{T}_A^{O|O'})^2/tr(\rho_A^0)^2$ we need the conformal block for the operator, which are given in \cite{Dotsenko:1984ca} \cite{Dotsenko:1985fp},
\begin{equation}
G(\eta,\bar{\eta})=\lvert\eta\rvert^{\frac{p}{q}}\lvert1-\eta\rvert^{\frac{p}{q}}\cdot\left[\frac{\sin{\left(\frac{\pi p}{q}\right)}\sin{\left(\frac{3\pi p}{q}\right)}}{\sin{\left(\frac{2\pi p}{q}\right)}}\lvert I_1(\eta)\rvert^2+\frac{\sin{\left(\frac{\pi p}{q}\right)}\sin{\left(\frac{\pi p}{q}\right)}}{\sin{\left(\frac{2\pi p}{q}\right)}}\lvert I_2(\eta)\rvert^2\right],
\end{equation}
\noindent where $I_1,I_2$ are defined as
\begin{align}
I_1(\eta)=\frac{\Gamma\left(\frac{3p}{q}-1\right)\Gamma\left(1-\frac{p}{q}\right)}{\Gamma\left(\frac{2p}{q}\right)}\cdot_2F_1\left(\frac{p}{q},\frac{3p}{q}-1,\frac{2p}{q},\eta\right),\\
I_2(\eta)=\eta^{1-\frac{2p}{q}}\frac{\Gamma\left(1-\frac{p}{q}\right)\Gamma\left(1-\frac{p}{q}\right)}{\Gamma\left(2-\frac{2p}{q}\right)}\cdot_2F_1\left(\frac{p}{q},1-\frac{p}{q},2-\frac{2p}{q},\eta\right),
\end{align}

\begin{figure}[H]
\centering
\subfloat[Case \Rmnum{1}]{
		\includegraphics[scale=0.32]{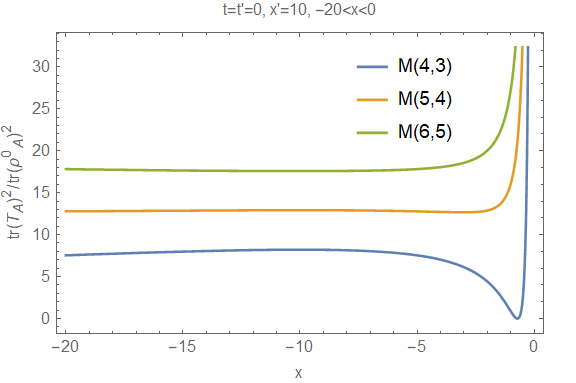}}
\subfloat[Case \Rmnum{2}]{
		\includegraphics[scale=0.37]{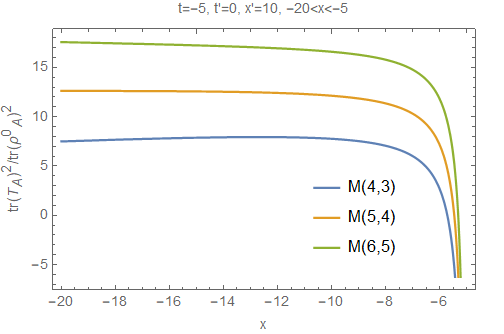}}
\\
\subfloat[Case \Rmnum{3}]{
		\includegraphics[scale=0.32]{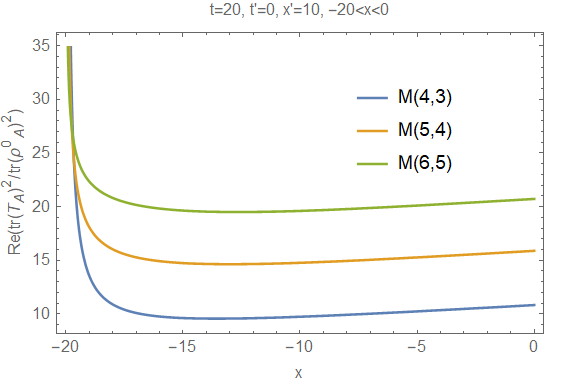}}
\subfloat[Case \Rmnum{3}]{
		\includegraphics[scale=0.32]{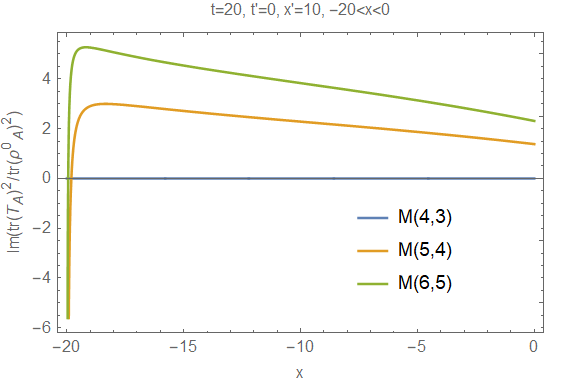}}
\caption{The plots of the logarithmic part of $\Delta S^{(2)}$ for the operator $\phi_{(2,1)}$ in minimal model $\mathcal{M}(4,3)$, $\mathcal{M}(5,4)$ and $\mathcal{M}(6,5)$. For all the plots $(t',x')$ is fixed to be $(0,10)$. (a) is for Case I, $t=0$, $x\in (-20,0)$. (b) is for Case II, $t=-10$, $x\in(-20,-10)$. (c) and (d) are for Case III, $t=20$, $x\in(-20,0)$.}
\label{fig:M(p,q),n=2}
\end{figure}

Specifically, we consider the operators $\phi_{(2,1)}$ in critical Ising model $\mathcal{M}(4,3)$, tricritical Ising model $\mathcal{M}(5,4)$ and three-state Potts at criticality $\mathcal{M}(6,5)$. We plot the results in Figure \ref{fig:M(p,q),n=2}. The results are similar with previous examples. 

\subsection{Summary of the examples}
In the above examples the second pseudo R\'enyi entropy or $tr (\mathcal{T}_A^{O|O'})^2/tr(\rho_A^0)^2$ show some general properties. For case I and case II all the examples support that $tr (\mathcal{T}_A^{O|O'})^2/tr(\rho_A^0)^2$ is real, but may be negative. For case III $tr (\mathcal{T}_A^{O|O'})^2/tr(\rho_A^0)^2$ may be complex or real, which depends on the theory and the operators.

For all the examples we find logarithmic part of the pseudo R\'enyi entropy would be divergent near the lightcone.  Similar behaviors have been found and discussed in \cite{Guo:2022sfl}. In the following sections we will show the divergence is universal, which only depends on the conformal dimension of the operator.

The results are shown in the following table.
\begin{table}[htb]   
\begin{center}    
\begin{tabular}{|c|c|c|c|}   
\hline   \ & $\partial\phi\bar{\partial}\bar \phi$ (n=2,3,4) & $\mathcal{V}_\alpha$ (n=2,3) & Minimal model $\phi_{(2,1)}$ (n=2)\\
\hline   \textbf{Case \Rmnum{1}} &  Positive & Real& Real\\   
\hline   \textbf{Case \Rmnum{2}}  &  Positive & Real& Real\\ 
\hline   \textbf{Case \Rmnum{3}} &  Positive & Complex&Complex\\     
\hline   
\end{tabular} 
\caption{Summary of the results for $tr (\mathcal{T}^{O|O'}_{A})^n/tr(\rho_A^0)^n$}  
\label{table:1}
\end{center}   
\end{table}

\subsection{General argument}
\subsubsection{The second pseudo R\'enyi entropy}
The second pseudo R\'enyi entropy is associated with conformal block (\ref{S2}). Using the cross ratios in Appendix.\ref{crossratiosection},  for case \Rmnum{1} and case \Rmnum{2} we have
\bea
1-\eta =\eta^*,\quad 1-\bar \eta =\bar \eta^* ,
\eea
where $*$ means the complex conjugation.
By using the cross symmetry we have
\bea
G(\eta,\bar \eta)=G(1-\eta,1-\bar \eta)=G(\eta^*,\bar \eta^*)=G(\eta,\bar \eta)^*
\eea
Further using (\ref{S2}) we find $tr (\mathcal{T}_A^{O|O'})^2/tr(\rho_A^0)^2$ should be real. 

However, for case III, $\eta$ is a real number, $\bar \eta$ is complex.
The conformal block can be expanded as $G(\eta,\bar \eta)\sim C_p F_p(\eta) \bar F(\bar \eta)$ with $C_p$ being real.  Using (\ref{S2}) we find generally $tr (\mathcal{T}_A^{O|O'})^2/tr(\rho_A^0)^2$ should be complex. This suggests that the spectra of the reduced transition matrix $\mathcal{T}_A^{O|O'}$ should include complex eigenvalues.

\subsubsection{n-th pseudo R\'enyi entropy for case \Rmnum{1}}
\begin{figure}[H]
\centering
\includegraphics[scale=0.4]{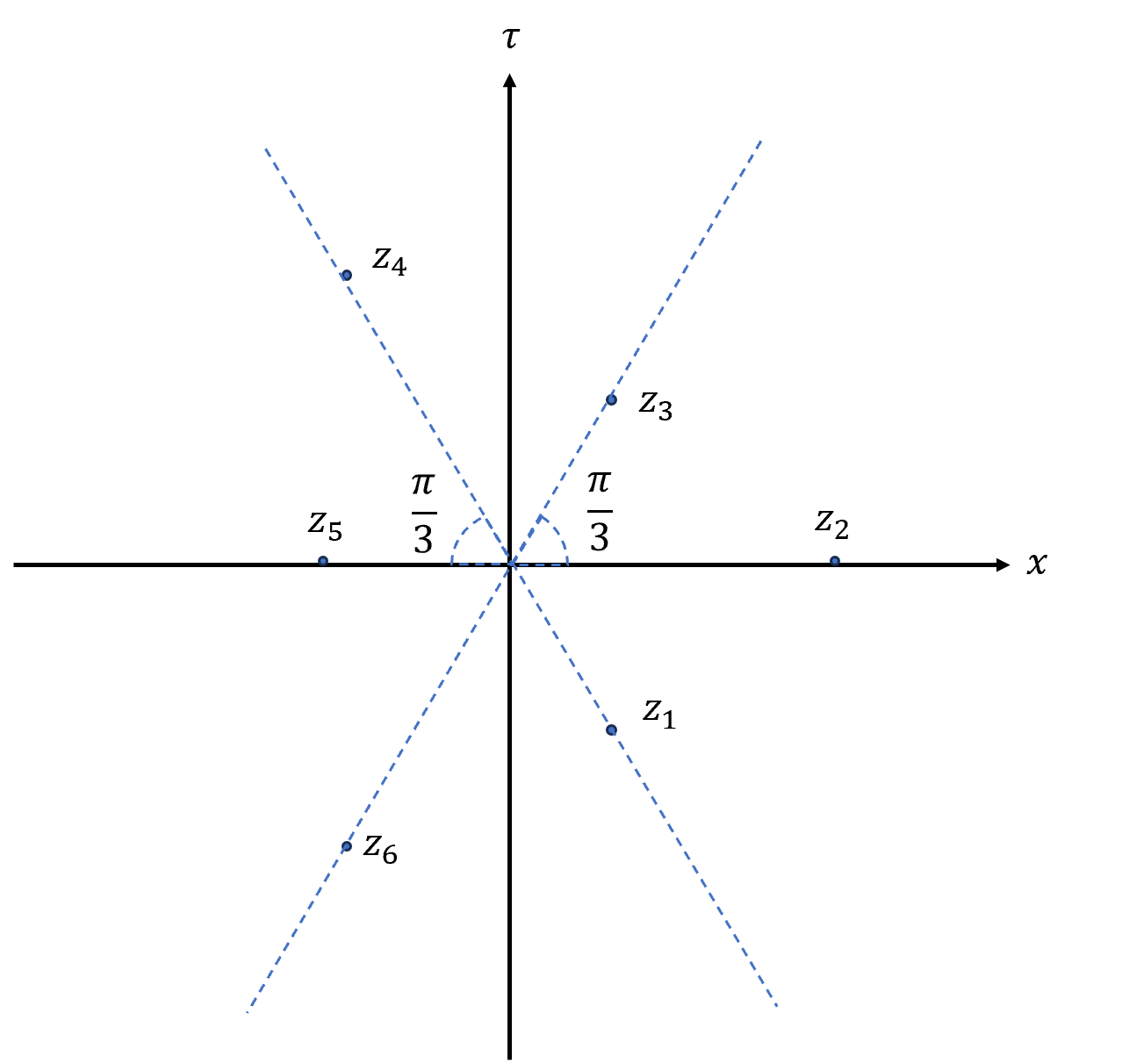}
\caption{Illustration of the coordinates of the operators $O(z_i,\bar z_i)$ ($i=1,...,6$) on the z-plane.}
\label{fig:6points}
\end{figure}

One could also obtain the above results by considering the correlation functions. According to \ref{trTncorrelator} and \ref{mapwtoz} 
\bea\label{trrhoonz}
tr (\mathcal{T}_{E,A})^n/tr(\rho_A^0)^n= \frac{\prod_{j=1}^{2n}(z_j \bar z_j)^{(1-n)h}}{n^{4nh}}  \frac{\langle O(z_1,\bar z_1)O(z_2,\bar z_2)...O(z_{2n-1},\bar z_{2n-1})O(z_{2n},\bar z_{2n})\rangle}{\langle O(w_1,\bar w_1)O(w_2,\bar w_2)\rangle_{\Sigma_1}^n},
\eea
where $z_j =w_j^{1/n}$ and $\bar z_j =\bar w_j^{1/n}$. To simplify the notations let us focus on $n=3$ and case \Rmnum{1}. The corresponding coordinates are
\bea
&&z_1=(-x)^{1/3} e^{-i\frac{\pi}{3}},\quad z_2= (x')^{1/3},\quad z_3=z_1 e^{i\frac{2\pi}{3}},\nn \\
&&z_4=z_2 e^{i\frac{2\pi}{3}},\quad z_5= z_1 e^{i\frac{4\pi}{3}},\quad z_6=z_2 e^{i\frac{4\pi}{3}},
\eea
as shown in Fig.\ref{fig:6points}. By directly calculations one can show that the coefficients $\frac{\prod_{j=1}^{2n}(z_j \bar z_j)^{1-n}}{n^{4n}}$ and the two point correlation functions $\langle O(w_1,\bar w_1)O(w_2,\bar w_2)\rangle_{\Sigma_1}$ are real.  The six point correlation functions can be written as
\bea
&&\langle O(z_1,\bar z_1) O(z_2,\bar z_2) O(z_3,\bar z_3)O(z_4,\bar z_4)O(z_5,\bar z_5)O(z_6,\bar z_6)\rangle \nn \\
&&=\langle \Psi| O(z_2,\bar z_2)O(z_5,\bar z_5) |\Psi\rangle ,
\eea
where we have defined the state $|\Psi\rangle:=  O(z_3,\bar z_3)O(z_4,\bar z_4)|0\rangle$. It is obvious that $\langle \Psi| =|\Psi\rangle^{\dagger}=\langle 0| O(z_1,\bar z_1)O(z_6,\bar z_6)$. Note that $O(z_2,\bar z_2)$ and $O(z_5,\bar z_5)$ are located on the time slice $\tau=0$. Thus we have
\bea
\langle \Psi| O(z_2,\bar z_2)O(z_5,\bar z_5) |\Psi\rangle^* =\langle \Psi| O(z_5,\bar z_5)O(z_2,\bar z_2) |\Psi\rangle=\langle \Psi|O(z_2,\bar z_2) O(z_5,\bar z_5) |\Psi\rangle,
\eea
where we have used the fact that $O(z_2,\bar z_2)$ commutes with $O(z_5,\bar z_5)$. Therefore, the six point correlation function is real.  It is not hard to generalize the result to $tr (\mathcal{T}_{E,A})^n/tr(\rho_A^0)^n$.
%, see the appendix for more details. 
It appears that the arguments presented above do not apply to case \Rmnum{2}. In the context of case \Rmnum{1}, the $2n$-point correlation functions can be mapped to the Euclidean z-plane. However, in case \Rmnum{2}, these correlation functions must undergo analytic continuation into real time. This fundamental distinction sets case \Rmnum{2} apart from case \Rmnum{1}, leading to a significantly different scenario. Addressing this issue requires further research, and we defer its resolution to future studies.

\section{Pseudo R\'enyi entropy near lightcone}\label{sectionlightcone}
\subsection{The second pseudo R\'enyi entropy near lightcone}
In all the examples we find the pseudo R\'enyi entropy is divergent near the lightcone.  In this section we would like to show the divergence of pseudo R\'enyi entropy is universal near the lightcone. Define the null coordinate $u=t+x$ and $v=t-x$. \\
Consider case \Rmnum{2}. For $u\to 0^-$,  the cross ratio $\eta$ is divergent 
\bea
\eta \to -\frac{i}{4\sqrt{x'}}(-u)^{-1/2},
\eea
while $\bar \eta$ is finite. The cross symmetry of conformal block is
\bea
G(\eta,\bar \eta)= \eta^{-2h}{\bar \eta}^{-2\bar h}G(\frac{1}{\eta},\frac{1}{\bar \eta}).
\eea
In general, conformal block $G(z,\bar z)$ can be expanded as
\bea
G(z,\bar z)=\sum_p C_p \mathcal{F}_p(z) \bar{\mathcal{F}}_p(\bar z),
\eea
where $C_p$ is the coupling constant, $\mathcal{F}(z)$ and $\bar {\mathcal{F}}(\bar z)$ are holomorphic and anti-holomorphic parts of the conformal block. Further, the conformal block can be expanded as a power series in $z$: $\mathcal{F}_p(z)=z^{h_p-2 h}\sum_{K=0}^{\infty} \mathcal{F}_{K} z^{K}$. 

Using the above results one can see in the lightcone limit $t+x\to 0^-$ we have 
\bea
G(\eta,\bar \eta)\simeq { \bar \eta}^{-2 \bar h}\bar{\mathcal{F}}_0(\frac{1}{\bar \eta})+O(\eta^{-1}),
\eea
Using (\ref{S2}) in the lightcone limit $u\to 0^-$ we have
\bea\label{caseIIlimit}
\frac{tr (\mathcal{T}_A^{O|O'})^2}{tr(\rho_A^0)^2}\to  (\eta (1-\eta))^{2h} (\bar \eta (1-\bar \eta))^{2\bar h}\bar{\mathcal{F}}_0(\frac{1}{\bar \eta})\sim (-u)^{-2h}.
\eea
Therefore, the leading contribution to $\Delta S^{(2)}$ is divergent as
\bea\label{lightcones2caseII}
\Delta S^{(2)}\to 2h\log |u|.
\eea
The result is only related to the conformal dimension $h$ of the operator.\\
Similarly, we can see the lightcone limit in case \Rmnum{3}, that is $u \to 0^+$. The cross ratio is also divergent
\bea
\eta \to -\frac{1}{4 \sqrt{x' }}u^{-1/2}.
\eea
Using same argument as case \Rmnum{2}, we can obtain 
\bea
\frac{tr (\mathcal{T}_A^{O|O'})^2}{tr(\rho_A^0)^2}\sim  (-u)^{2h}=e^{-i 2\pi h} u^{-2h}.
\eea
Thus in the lightcone limit $u\to 0^+$, $\Delta S^{(2)}$ is also divergent as
\bea\label{lightcones2caseIII}
\Delta S^{(2)}\to 2h\log u.
\eea
The pseudo R\'enyi entropy is also only related to the conformal dimension of the operator.

Near the lightcone $u\sim 0$, the real part of the second pseudo R\'enyi entropy is divergent. Both in case \Rmnum{2} and case \Rmnum{3} the divergent part is associated with the conformal dimension $h$ of the operator. The imaginary part of the second pseudo R\'enyi entropy is non-universal, which depends on the anti-holomorphic cross ratio $\bar \eta$. In case \Rmnum{2} we have shown $\frac{tr (\mathcal{T}_A^{O|O'})^2}{tr(\rho_A^0)^2}$ must be real. Thus the imaginary part of the second pseudo R\'enyi entropy should be $i\pi$ or $0$, which depends on the conformal block of anti-holomorphic part as we can see from (\ref{caseIIlimit}).

\subsection{$n$-th pseudo R\'enyi entropy near lightcone}
The result in last section can be generalized to the $n$-th pseudo R\'enyi entropy. To map the manifold $\Sigma_n$ to  $z$ complex plane, we use the coordinate transformation $z=w^{1/n}$. The corresponding coordinates on $z$-plane are given by
\bea
&&z_1=w_1^{1/n}=(x+t- i\epsilon)^{1/n}, \quad \bar z_1={\bar w_1}^{1/n}=(x-t+i\epsilon)^{1/n}\nn \\
&&z_2=w_2^{1/n}=(x'+i\epsilon)^{1/n},\quad \bar z_2={\bar w_2}^{1/n}=(x'-i\epsilon)^{1/n},\nn \\
&&z_3= z_1 e^{\frac{2\pi i}{n}},..., \bar z_3=\bar z_1 e^{-\frac{2\pi i}{n}}, \nn \\
&&z_{2n-1}=z_1 e^{(n-1)\frac{2\pi i}{n}},..., \ \bar z_{2n-1}=\bar z_1 e^{-(n-1)\frac{2\pi i}{n}},\\
&&z_{2n}=z_2 e^{(n-1)\frac{2\pi i}{n}},...,\ \bar z_{2n}=\bar z_2 e^{-(n-1)\frac{2\pi i}{n}}.
\eea 

\noindent In case II near the lightcone $u\to 0^{-}$, we have
\bea\label{lightconelimitz}
z_1\simeq e^{-\frac{i\pi}{n}}(-u)^{1/n},\quad z_3=\simeq e^{\frac{i\pi}{n}}(-u)^{1/n}.
\eea 
That is $z_1 -z_3 \to 0$ in the limit $u\to 0^{-}$.Similarly, we have $z_{2j-1}-z_{2j+1}\to 0$ with $j={1,2,...,n-1}$ . 

From the expression of $tr (\mathcal{T}_{E,A})^n/tr(\rho_A^0)^n$ (\ref{trrhoonz}) we can see that there are two sources for the divergence in the lightcone limit $u\to 0^-$. One is the coefficients $\prod_{i=1}^{n}z_{2i-1}^{(1-n)h}$ since $z_{2i-1}\to 0$. The other one is from the $2n$-point correlations functions, which can be determined by using operator product expansion (OPE), which has the following form:
\bea
O(z,\bar z)O(z',\bar z')=\sum_p \sum_{K,\bar K} C_p^{\{K,\bar K\}}(z-z')^{h_p+K-2h}(\bar z-\bar z')^{\bar h_p+\bar K-2h} O_{p}^{K,\bar K}(z',\bar z'),
\eea
where $h$ is the conformal dimension of the operator $O$,  $p$ labels the operators that appear in the OPE, $K,\bar K$ labels the descendants,  $ C_p^{\{K,\bar K\}}$ are the coupling constants.

Let us consider the $n=2$ as an example. The coefficients $\prod_{i=1}^{2}z_{2i-1}^{-h}\sim (-u)^{-h}$ by using (\ref{lightconelimitz}).  In the lightcone limit we have $z_1-z_3\sim (-u)^{1/2}$, thus the OPE can be approximated by
\bea
O(z_{1},\bar z_{1})O(z_{3},\bar z_{3})\simeq (z_{1}- z_{3})^{-2h}\sum_{\bar K}(\bar z_{1}-\bar z_{3})^{ \bar K-2h} O_{0}^{0,\bar K}(z_{3},\bar z_{3}),
\eea
where $O_0$ denotes the identity operator. The leading contribution comes from the identity operator and its its anti-holomorphic descendants, such as $\bar T$. Using the fact $z_1-z_3\sim (-u)^{1/2}$, the 4-point correlation function has the limit
\bea\label{3point}
\langle O(z_{1},\bar z_{1})O(z_{2},\bar z_{2})O(z_{3},\bar z_{3})O(z_{4},\bar z_{4})\rangle \sim (-u)^{-h} \langle \mathcal{O}^{\bar K}(z_3,\bar z_3)  O(z_{2},\bar z_{2})O(z_{4},\bar z_{4})\rangle.
\eea
where 
\bea
\mathcal{O}^{\bar K}(z_3,\bar z_3):=\sum_{\bar K}(\bar z_1-\bar z_{3})^{ \bar K-2h} O_{0}^{0,\bar K}(z_{3},\bar z_{3}),
\eea
which can be taken as summations of the anti-holomorphic descendants. The $3$-point correlation function in (\ref{3point}) is expect to be finite, thus we find 
\bea
tr (\mathcal{T}_{A})^2/tr(\rho_A^0)^2\sim (-u)^{-2h},
\eea
or equally $\Delta S^{(2)}\sim 2h \log (-u)$, which is consistent with the discussion by using conformal block (\ref{caseIIlimit}).
We can as well utilize Operator Product Expansion (OPE) to simplify the $2n$-point correlation functions mentioned in (\ref{trrhoonz}) as the lightcone limit $u\to 0^-$. However, it is crucial to note that the diverge behavior significantly based on the details of the theory. 

Let us go on considering the case with $n=3$.  In the limit $u\to 0^-$ we have $z_1,z_3,z_5 \sim (-u)^{1/3}$, thus the coefficients  $\prod_{i=1}^{3}z_{2i-1}^{-2h}\sim (-u)^{-2h}$. The 6-point correlation function is
\bea
\langle O(z_1,\bar z_1)O(z_2,\bar z_2)O(z_3,\bar z_3)O(z_4,\bar z_4)O(z_5,\bar z_5)O(z_6,\bar z_6)\rangle,
\eea
which is also divergent.  Since $z_1-z_3\sim (-u)^{1/3}$ we have the OPE
\bea
O(z_1,\bar z_1)O(z_3,\bar z_3)=\sum_p \sum_{K,\bar K} C_p^{\{K,\bar K\}}(z_1-z_3)^{h_p+K-2h}(\bar z_1-\bar z_3)^{\bar h_p+\bar K-2h} O_{p}^{K,\bar K}(z_3,\bar z_3).
\eea
In the $n=2$ case, we only consider contributions from anti-holomorphic descendants of the identity operator. However, it is essential to take into account other operators, such as the stress-energy tensor $T(z)$, as these operators in the OPE have strong correlations with $O(z_5, \bar{z}_5)$. This complicates the discussions significantly.   

Same as the case $n=2$, the identity operator and its anti-holomorphic descendants will give the contribution 
\bea\label{identitycontribution}
&&(z_1-z_3)^{-2h}(\bar z_1-\bar z_3)^{-2h+\bar K} \langle \mathcal{O}^{\bar K}(z_3,\bar z_3) O(z_2,\bar z_2)O(z_4,\bar z_4)O(z_5,\bar z_5)O(z_6,\bar z_6)\rangle\nn \\
&&\sim |u|^{-\frac{2h}{3}}.
\eea
Unlike the case $n=2$ the holomorphic descendants of identity operator will also contribute. Consider the stress energy tensor $T(z)$ as an example,  the OPE gives
\bea
&&(z_1-z_3)^{-2h+2}(\bar z_1-\bar z_3)^{-2h} \langle T(z_3)O(z_2,\bar z_2)O(z_4,\bar z_4)O(z_5,\bar z_5)O(z_6,\bar z_6)\rangle\nn \\
&&\sim (z_1-z_3)^{-2h+2} (z_3-z_5)^{-2}(\bar z_1-\bar z_3)^{-2h} \langle O(z_2,\bar z_2)O(z_4,\bar z_4)O(z_5,\bar z_5)O(z_6,\bar z_6)\rangle\nn \\
&& \sim |u|^{-\frac{2h}{3}},
\eea
where in the second step we use the Ward identities and only keep the leading contributions.  Similar calculations can be done for other descendants.  It can shown the results are all divergent as $|u|^{-\frac{2h}{3}}$. 

Further, we should consider the possible contributions from other primary fields and their descendants. Take the primary operator $O_p$ as an example. The contribution is given by
\bea\label{sixpOOPE}
(z_1-z_3)^{-2h+h_p}(\bar z_1-\bar z_3)^{-2h+h_p} \langle O_p(z_3,\bar z_3)O(z_2,\bar z_2)O(z_4,\bar z_4)O(z_5,\bar z_5)O(z_6,\bar z_6)\rangle.
\eea
Note that $z_3-z_5\sim |u|^{1/3}$. We can further expand the product $O_p(z_3,\bar z_3)O(z_5,\bar z_5)$. Assume the fusion rule
%If $O_p=O$, thus $h_p=h$, the leading contribution of the OPE for $O_p(z_3,\bar z_3)O(z_5,\bar z_5)$ would be $(z_3-z_5)^{-2h}\sim |u|^{-\frac{2h}{3}}$. Combining  with (\ref{sixpOOPE})  we can see the contribution from primary operator $O_p$ would be divergent as $|u|^{-h}$.  In this case the operator $O$ would appear in the OPE of $O O$, that is $O\times O= I+ O+...$.
%On the contrary, if $O$ doesn't appear in the OPE of $O\times  O= I + O_p+...$ with $O_p \ne O$.  We would have 
$O_p \times O = O_{p'}+...$ with $h_{p'}\ge 0$. The leading contribution of the OPE of $O_p(z_3,\bar z_3)O(z_5,\bar z_5)$ would be $(z_3-z_5)^{-h-h_p+h_{p'}}$ with $h_{p'}\ge 0$. Combining with (\ref{sixpOOPE}) we find the leading divergent term of 6-point correlation function is  $|u|^{-h+\frac{h_{p'}}{3}}$. Recall that the contribution from the identity and its anti-holomorphic descendants is divergent as $|u|^{-\frac{2h}{3}}$  (\ref{identitycontribution}). If $h_{p'}\ge h$, we have $-h+\frac{h_{p'}}{3}\ge -\frac{2h}{3}$,  the 6-point correlation function is divergent as $|u|^{-\frac{2h}{3}}$. Combining with the divergent term from the coefficients $\prod_{i=1}^{3}z_{2i-1}^{-2h}\sim (-u)^{-2h}$, we have
\bea\label{s3universal}
\Delta S^{(3)} \sim \frac{4 h}{3}\log|u|.
\eea
Otherwise, if $h_{p'}< h$, we would have $\Delta S^{(3)} \sim \frac{1}{2}(3h-\frac{h_{p'}}{3})\log |u|$. 

Although the above argument can be extended to any arbitrary $n$, the results will also become more complicated. For the general case of $n$, we do not expect a simple conclusion. Instead, it depends on more specific details of the theory. 

\subsection{Examples for pseduo R\'enyi entropy near lightcone}\label{lightconeexamples}
As we show in last section, near the lightcone, the pseudo R\'enyi entropy should be divergent  as $\log|u|$ (\ref{lightcones2caseII})(\ref{lightcones2caseIII}). In this section we would like use examples to check the results. For the second pseudo R\'enyi entropy the result is universal. We show that the divergent term is $2h \log|u|$ for both case \Rmnum{2} and \Rmnum{3}, which only depends on the conformal dimension of the operator. While for $n\ge 3$ the result would be more complicated, it depends on the details of theory, that is the OPE of the operators in the theory.

\par
\begin{figure}[H]
\centering
\subfloat[$\partial\phi\bar{\partial}\phi$ in case \Rmnum{2}]{
		\includegraphics[scale=0.4]{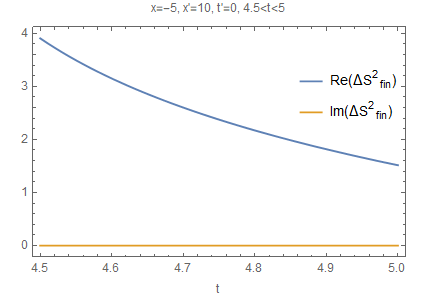}}
\subfloat[$\partial\phi\bar{\partial}\phi$ in case \Rmnum{3}]{
		\includegraphics[scale=0.41]{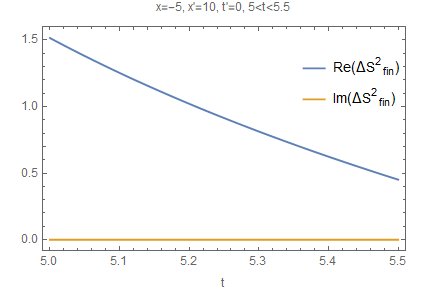}}
\\
\subfloat[$\mathcal{V}_{\alpha}$ in case \Rmnum{2}]{
		\includegraphics[scale=0.38]{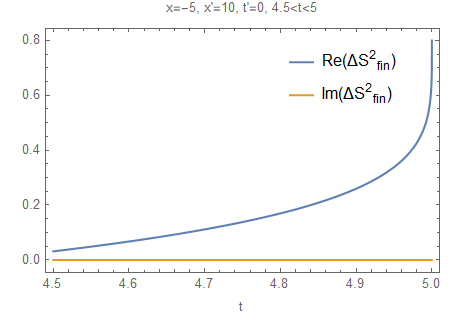}}
\subfloat[$\mathcal{V}_\alpha$ in case \Rmnum{3}]{
		\includegraphics[scale=0.38]{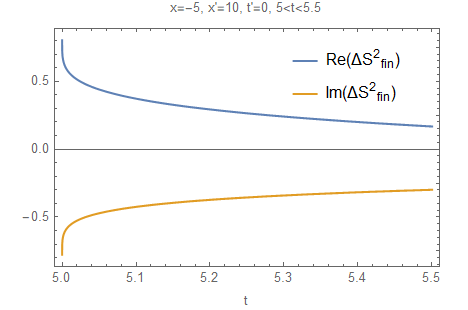}}
\\
\subfloat[$\phi_{(2,1)}$ in case \Rmnum{2}]{
		\includegraphics[scale=0.38]{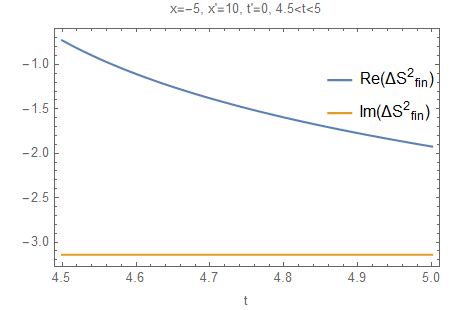}}
\subfloat[$\phi_{(2,1)}$ in case \Rmnum{3}]{
		\includegraphics[scale=0.38]{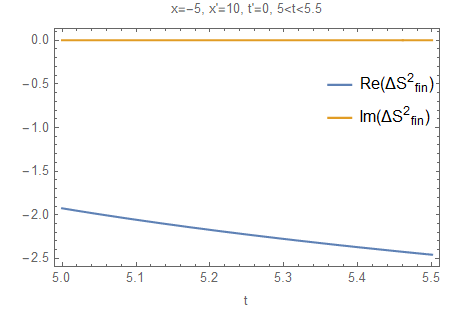}}
\caption{The plots of $\Delta S^{(2)}_{\text{fin}}$ for the operators $\partial\phi\bar{\partial}\phi$, $\mathcal{V}_{\alpha=\frac{1}{2}}$ and $\phi_{(2,1)}$ in minimal model $\mathcal{M}(4,3)$. For all the plots $(t',x')$ is fixed to be $(0,10)$, (a)(c)(e) is for Case II, $x=-5$, $t\in [4.5,5]$; (b)(d)(f) is for Case III, $x=-5$, $t\in[5,5.5]$.}
\label{fig:S_fin}
\end{figure}

 Firstly, consider the operator $\p \phi \bar{\p} \phi$. The second pseudo R\'enyi entropy can be obtained by using (\ref{R2phicaseII}) for case \Rmnum{2} and case \Rmnum{3}. It is easy to check the leading divergent term is $\Delta S^{(2)}\sim 2\log (-u)$, which is consistent with result (\ref{lightcones2caseII})(\ref{lightcones2caseIII}).  For the vertex operator $\mathcal{V}_\alpha$ and $\phi_{(2,1)}$ in the Minimal Model one could also check this directly by using the expressions in section.\ref{vertexoperatorsection} and section.\ref{minimalmodelsection}. 
It is also helpful to define the difference between the pseudo R\'enyi entropy and the universal divergent term, that is 
\bea
\Delta S^{(2)}_{\text{fin}}:= \Delta S^{(2)}-2 h \log|u|.
\eea
We show $\Delta S^{(2)}_{\text{fin}}$ for the three examples in Fig.\ref{fig:S_fin}.

For $n\ge 3$ the results would be more subtle.  Consider the vertex operator $\mathcal{V}_\alpha$, one could obtain $\Delta S^{(3)}$ by calculating the 6-point correlation functions. Using the notation in \cite{Nakata:2020luh} the result is
\bea\label{s3v}
\Delta S^{(3)}=\frac{1}{2}\log \frac{4}{1+3(|\eta^{32}_{14}|^{8\alpha^2}+|\eta^{14}_{56}|^{{8\alpha^2}}+|\eta^{56}_{32}|^{{8\alpha^2}}},
\eea
where $\eta^{ij}_{mn}$ are the cross ratios defined as
\bea
\eta^{ij}_{mn}=\frac{(z_m-z_n)(z_i-z_j)}{(z_m-z_i)(z_n-z_j)},
\eea
where $1\le i,j,m,n\le 6$. In the lightcone limit $u\to 0^-$, the cross ratios would be divergent as
\bea
\eta^{32}_{14},\eta^{14}_{56},\eta^{56}_{32} \sim |u|^{-\frac{1}{3}}.
\eea
Using the result (\ref{s3v}) we conclude that 
\bea
\Delta S^{(3)}\sim \frac{2\alpha^2}{3}\log |u|,
\eea
which is consistent with (\ref{s3universal}) with $h=\frac{\alpha^2}{2}$. In last section we have shown the result (\ref{s3universal}) is based on the assumption that $h_{p'}\ge h$.

%The finite terms are non-universal, which depends on details of the operators and the theory. Let us consider the operator $\p \phi \bar{\p} \phi$, $\mathcal{V}_\alpha$ in case \Rmnum{2}, by using [\ref{}], [\ref{}], and the limitation of cross ratio in case \Rmnum{2},
%\begin{align}
%\eta&=\frac{1}{2}-i\frac{x+t+x'}{4\sqrt{-x-t}\sqrt{x'}}=\frac{1}{2}-i\frac{u+x'}{4\sqrt{-u}\sqrt{x'}}
%\end{align}
%we have
%\begin{align}
%\Delta S^{(2)}_{\text{fin}}(\p \phi \bar{\p} \phi)=-\log{\left[{(64x')}^{-2}(1+\bar{\eta}(\bar{\eta}-1))^2(48x'u+4(x'+u)^2)^2\right]}
%\end{align}
%we notice that $\bar{\eta}\stackrel{u\rightarrow 0}\longrightarrow C$, which is a constant.
%\begin{equation}
%\lim_{u\rightarrow 0}\Delta S^{(2)}_{\text{fin}}(\p \phi \bar{\p} \phi)=-\log{\left[\left(\frac{x'}{16}\right)^2(1+C(C-1))^2\right]}
%\end{equation}
%As we can see, it is finite near the lightcone.

%For $n=3,4$ we also have the expressions of $tr (\mathcal{T}_A^{O|O'})^n/tr(\rho_A^0)^n$ \ref{R3phicaseII} and \ref{R4phicaseII} for case \Rmnum{2}. One can check that they are both divergent as $1/u^2$. Therefore, $\Delta S^{(n)}\sim \frac{2}{n-1}\log(-u)$ for $n=2,3$, which is consistent with \ref{divergentRenyin}. 
For the vertex operator  $\mathcal{V}_\alpha$ we have the OPE
\bea
\mathcal{V}_\alpha\times \mathcal{V}_\alpha =I+ \mathcal{V}_{2\alpha}.
\eea
The operator $\mathcal{V}_{2\alpha}$ may give the contribution to the divergent term of 6-point correlation function. 
Further, we have
\bea
\mathcal{V}_{2\alpha}\times \mathcal{V}_{\alpha}=\mathcal{V}_{\alpha}+\mathcal{V}_{3\alpha}.
\eea
Note that the conformal dimensions of  $\mathcal{V}_{\alpha}$ and $\mathcal{V}_{3\alpha}$ are $\frac{\alpha^2}{2}$ and $\frac{9\alpha^2}{2}$. These two operators actually correspond the operator $O_{p'}$ that we discuss in last section. It is obvious that the condition $h_{p'}\ge h$ is satisfied. Thus this example can be taken as a nice check of the result (\ref{s3universal}).

For the operator $\partial \phi \bar \partial \bar \phi$ we should be more careful. In the discussions of last section we implicitly assume the correlation functions cannot factor as holomorphic and anti-holomorphic parts, which is not correct for  $\partial \phi \bar \partial \bar \phi$. Near the lightcone the divergence of correlation function comes from the holomorphic field. Take $n=3$ as an example, the correlation function is proportional to
\bea
\langle \partial \phi(z_1)\partial \phi(z_2)\partial \phi(z_3)\partial \phi(z_4)\partial \phi(z_5)\partial \phi(z_6)\rangle.
\eea
Near the lightcone we know $z_1,z_2,z_3\sim |-u|^{1/3}$. The leading divergent term of the above correlator is given by
\bea
\langle \partial \phi(z_1)\partial \phi(z_3)\partial \phi(z_5)\rangle \langle \partial \phi(z_2)\partial \phi(z_4)\partial \phi(z_6)\rangle,
\eea
which is vanishing. Thus the final result is finite in the limit $|u|\to 0$.  The divergence of the pseudo R\'enyi entropy comes from the coefficients $\prod_{i=1}^{3}z_{2i-1}^{-2}\sim (-u)^{-2}$. Thus the third pseudo R\'enyi entropy should be divergent as $\log|u|$. 

For $n=4$ we can use similar argument as the case $n=3$, the leading divergent term of the 8-point correlation function is given by 
\bea
\langle \prod_{i=1}^{4}\partial \phi(z_{2i-1})\rangle \langle \prod_{j=1}^{4}\partial\phi(z_{2j})\rangle\sim |u|^{-1}.
\eea
Combining with the coefficients $\prod_{i=1}^{4}z_{2i-1}^{-3}\sim |u|^{-3}$, we have $\Delta S^{(4)}\sim \frac{4}{3} \log |u|$. In Appendix.\ref{appfreescalar} we check this by directly calculation using Wick theorem.

\section{Pseudo-Hermitian condition}\label{sectionpseudo}
In previous sections we study the pseudo R\'enyi entropy for the transition matrix (\ref{transitionlocal}). We mainly focus on three cases, where the location of the operator $O(t,x)$ is different. From several examples we find the results summarized in the Table.\ref{table:1}. For the cases \Rmnum{1} and case \Rmnum{2} the pseudo R\'enyi entropy is real for all the examples we consider. For case \Rmnum{3} the results are generally not real. The pseudo R\'enyi entropy being complex implies that some of the eigenvalues of the reduced transition matrix $\transA$ are complex. The pseudo R\'enyi entropy being real implies the eigenvalues of $\transA$ are real or complex coming in conjugated pairs. 

In \cite{Guo:2022jzs}  the authors point out one could understand the real-valued condition by using pseudo-Hermiticity. If an operator $M$ satisfies
\bea
M^\dagger =\eta M \eta^{-1},
\eea
where $\eta$ is an invertible and Hermitian operator, we say $M$ is $\eta$-pseudo-hermitian.  
A notable fact is that the diagonalizable operator $M$ is $\eta$-pseudo-hermitian operator $M$ if and only if the eigenvalues are real or complex coming in complex pairs. 

In \cite{Guo:2022jzs} it is found  the general $\eta$-pseudo-Hermitian transition matrix can be written as 
\bea
\mathcal{T}=\frac{|\psi\rangle \langle \psi| \eta}{\langle \psi| \eta |\psi\rangle}.
\eea
If $\eta=\eta_A\otimes \eta_{\bar A}$ with both $\eta_A$ and $\eta_{\bar A}$ being invertible and hermitian, one could show that the reduced transition matrix would be pseudo-Hermitian. Thus the eigenvalues would be real or complex coming in conjugated pairs. The pseudo R\'enyi entropy is expected to be real.  Further, if $\eta_{A}$ and $\eta_{\bar A}$ are positive or negative operator, the eigenvalues are expected to be positive thus the pseudo R\'enyi entropy would be positive. 
The result of our previous examples imply the transition matrix for case \Rmnum{1} and \Rmnum{2} may be pseudo-Hermitian. Our goal in this section is to investigate whether the reduced density matrix can be written as pseudo-Hermitian form.

\subsection{Translation and boost operators}
In this section we would like to introduce the smearing operators with stress energy tensor which are related to our problem. A local QFT has stress energy tensor $T_{\mu\nu}$ which is conversed $\partial^\mu T_{\mu\nu}=0$. One could construct the Hamiltonian 
\bea
H:= \int_{V} d^3 x T_{00},
\eea 
where $V$ denotes the whole region on the time slice $t=0$, which generates the time translation. The $i$-th component of momentum operator is 
\bea
P^i= \int_V d^3 x T^{i0},
\eea
where $i=x,y,z$, which generates the translation on the $i$-the direction. The boost in the $x$-direction is generated by the modular Hamiltonian 
\bea
K= \int_V d^3x xT_{00}.
\eea

For our motivation we would like to introduce the similar operators located in $A$ or $\bar A$, that is the operators 
\bea
T_{00}(f):= \int d^3x f(\vec{x})T_{00}(0,\vec{x}),\quad T_{0i}(g):= \int d^3x g(\vec{x})T_{0i}(0,\vec{x}),
\eea
where $f$ and $g$ are function supported in region $A$. Similarly, we can define the local operators in $\mathcal{R}(\bar A)$ by using the functions supported in $\bar A$. Specially, we are interested in the following ones,
\bea\label{localdefinefree}
&&H_A:= T_{00}(H(x)),\quad H_{\bar A}:= T_{00}(H(-x)),\nn \\
&&P_{A,i}:=T_{i0}(H(x)),\quad P_{\bar A,i}:=T_{i0}(H(-x)),\nn \\
&&K_A:=T_{00}(xH(x)),\quad K_{\bar A}=T_{00}(-xH(-x)).
\eea
It is well known that the generators $H,P^i,K$ should satisfy the Poincare algebra. We have\footnote{Here we will only use the translation and boost in the $x$-direction. Only related generators are listed.}  
\bea\label{Poincare}
[H,P_i]=0,\quad [K,P_x]=-i H,\quad  [K,H]=-i P_x.
\eea
By using the Baker–Campbell–Hausdorff (BCH) formula one could show
\bea
e^{-K\pi} P_x e^{K \pi}= -P_x,\quad e^{-K\pi} H e^{K \pi}= -H.
\eea
Define the operator $\eta:= e^{K \pi}e^{i P_x a}$ for the real parameter $a$. It can be shown $\eta$ is Hermitian,
\bea\label{etahermitian}
\left(e^{K \pi}e^{i P_x a}\right)^\dagger =e^{-i P_x a}e^{K \pi}=e^{K \pi}e^{i P_x a}.
\eea
Since $P_x$, $K$ and $H$ are all constructed by stress energy tensor, the Poincare algebra  should give some constraints on the commutators of stress energy tensor $[T^{\mu\nu},T^{\rho\sigma}]$. The form of the commutators can be determined up to the so-called Schwinger terms, which need to be total derivatives\cite{Deser:1967zzf}. Therefore, the commutators of $H_{A(\bar A)},P_{A(\bar A)}^i,K_{A(\bar A)}$ may be different from the Poincare algebra (\ref{Poincare}). In this paper we will mainly focus on 2-dimensional CFTs, for which the commutators of stress energy tensor are known. For free scalar theory we also calculate the commutators, which are shown in Appendix.\ref{appfreecomu}.
\subsection{2 dimensional CFTs}
For 2 dimensional CFTs the commutators of stress energy tensor is given by
\bea\label{stresscom}
[T_{uu}(u),T_{uu}(u')]=i \left(T_{uu}(u)+T_{uu}(u')\right)\p_u \delta(u-u')-\frac{ic}{24\pi }\p_u^3 \delta(u-u'),
\eea
where $u=t-x$. We have the similar commutator relation for $T_{vv}$ with $v=t+x$. Define the smearing operators $T_{00}(f):= \int dx f(x)T_{00}$, $T_{0x}(g):= \int dx f(x)T_{0x}$ with $T_{00}=T_{uu}+T_{vv}$ and $T_{0x}=T_{vv}-T_{uu}$. Similar as the definition (\ref{localdefinefree}) let us define the local operators 
\bea\label{localoperators2D}
&&P_{A,x}:=T_{0x}(H(x)),\quad P_{\bar A,x}:=T_{0x}(H(-x)),\nn \\
&&K_A:=T_{00}(xH(x)),\quad K_{\bar A}:=T_{00}(-xH(-x)).
\eea
One could evaluate the commutators of $H_{A(\bar A)},P_{A(\bar A)}^i,K_{A(\bar A)}$ by using (\ref{stresscom}), see the appendix for details.  The commutators are given by
\bea\label{localcom2d}
&&[K_A, H_A]=-i P_{A,x},\quad  [K_A, P_{A,x}]=-i H_A,\nn \\
&&[K_{\bar{A}}, H_{\bar A}]=i P_{\bar A,x},\quad  [K_{\bar A}, P_{\bar A,x}]=i H_{\bar A},
\eea 
and
\bea
[K_A,P_{\bar A,x}]=0,\quad [K_{\bar A},P_{A,x}]=0,\quad [K_A,K_{\bar A}]=0.
\eea
By using BCH formula we would have 
\bea
e^{\pi K_A} P_{A,x}e^{-\pi K_A}=-P_{A,x}.
\eea
Further it can be shown
\bea
[e^{\pi K_A}e^{i P_{A,x} a} ]^\dagger=(e^{\pi K_A}e^{i P_{A,x }a}e^{-\pi K_A})e^{\pi K_A}=e^{\pi K_A}e^{i P_{A,x} a} ,
\eea
which means the operator $\eta_A:=e^{\pi K_A} e^{i P_{A,x} a} $ is a Hermitian operator. 
%Further, it can be shown that
%\bea
%e^{\pi K_A}e^{i P_{A,x} a} =e^{-\frac{i }{2}P_{A,x} a}e^{\frac{\pi }{2}K_A} \left( e^{-\frac{i }{2}%P_{A,x} a}e^{\frac{\pi }{2}K_A}\right)^\dagger,
%\eea
%which means the operators $\eta_A$ is non-negative operators. Similarly, we can define the operator $\eta_{\bar A}:=e^{\pi K_{\bar A}}e^{i P_{\bar A,x} a} $, which is also non-negative operator.

\subsection{Pseudo Hermitian }

\begin{figure}[h]
\centering
\includegraphics[width=7cm]{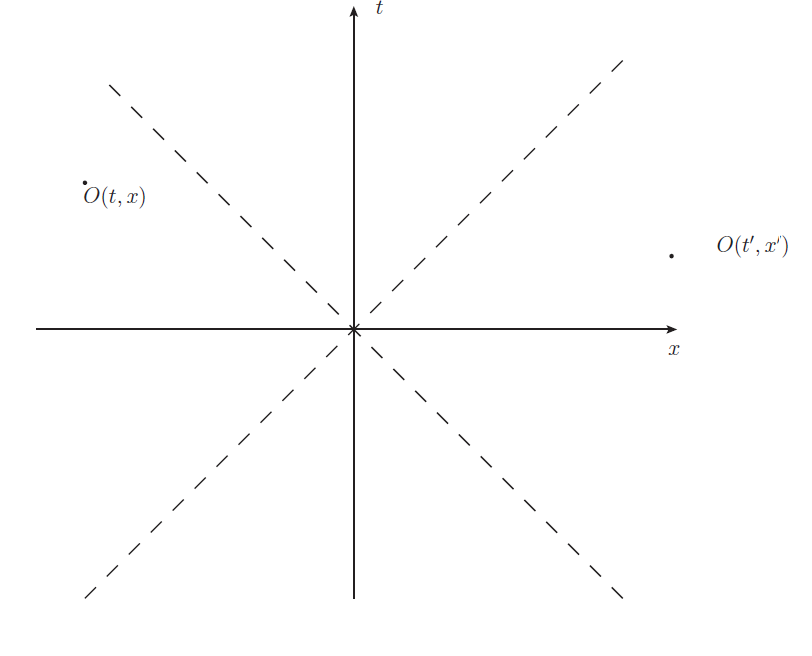}
\caption{The case that operators $O(t,x)$ and $O(t',x')$ are located in left and right Rindler wedge.}\label{p1}
\end{figure}

Let us consider the transition matrix (\ref{transitionlocal}) with $O(t,x)$  and $O'(t,x)$ located in left and right Rindler wedges, respectively as shown in Fig.\ref{p1}.
We would like to show the transition matrix (\ref{transitionlocal}) is $\eta$-pseudo-Hermitian with the form $\eta=\eta_A\otimes \eta_{\bar A}$. \\
Firstly, let us consider the case \Rmnum{1}. Note that we have $x<0$ and $x'>0$. Using the translation and boost operator, we have
\bea
e^{i P_x(x'+x)}e^{-K \pi}O(0,x)e^{K \pi}e^{-i P_x(x'+x)}=e^{i P_x(x'+x)}O(0,-x)e^{-i P_x(x'+x)}=O(0,x').
\eea
The transition matrix can be written as
\bea
O(0,x)|0\rangle \langle 0| O(0,x')=O(0,x)|0\rangle \langle 0| O(0,x)e^{K \pi}e^{-i P_x(x'+x)},
\eea
by using the fact that $K|0\rangle=0$ and $P_x|0\rangle=0$. The operator $\eta=e^{K \pi}e^{-i P_x(x'+x)}$ is Hermtian and invertible (\ref{etahermitian}). Thus the transition matrix (\ref{transitionlocal}) in  this case is pseudo Hermitian. In fact we will further show $\eta$ can be written as $\eta_A\otimes \eta_{\bar A}$ where both $\eta_A$ and $\eta_{\bar A}$ are Hermitian and invertible.

In last section we define the local operators $K_{A(\bar A)}$ and $P_{x,A(\bar A)}$ by choosing the smearing functions $f$ and $g$ for the operators $T_{00}(f)$ and $T_{0x}(g)$ (\ref{localoperators2D}).  Abviously, we have the relation $P_x=P_{x,A}+P_{x,\bar A}$ and $K=K_A+K_{\bar A}$ by definitions.
There is a subtle point for the operators $P_{x,A(\bar A)}$. If we  calculate the commutator $[P_{x,A}, P_{x,\bar A}]$ by using (\ref{localoperators2D}) and (\ref{appendixcftstress}), a boundary term located at the entanglement boundary $\p A$ will appear. Therefore, it seems we cannot decompose $e^{i P_{x}}=e^{i P_{x,A}}e^{i P_{x,\bar A}}$.
However, we will argue that the boundary will not appear if we carefully consider the process to evaluate pseudo R\'enyi entropy. 

It is well known that the (pseudo) Rényi entropy exhibits UV divergence in quantum field theories (QFTs) and requires regularization. As shown in \cite{Cardy:2016fqc} the regularization can be taken as projection $P_{\p A}^\epsilon$ in the Hilbert space, which removes small spatial region of thickness $\epsilon$ around the entanglement boundary $\p A$. For a pure state $|\phi\rangle $ that is to say we would consider the regularized states
\bea
|\phi\rangle_\epsilon:=P_{\p A}^\epsilon |\phi\rangle.
\eea
In the Euclidean path integral formulation the projection can be understood as introducing a small slit with lengh $\epsilon$ around the boundary  $\p A$. Using this the authors in \cite{Cardy:2016fqc} derive the modular Hamiltonian and R\'enyi entropy for a lot of known examples. With considering the regularization   the modular Hamiltonian of the subsystem $A=(0,+\infty)$ should be $K_A=\int_\epsilon^\infty dx x T_{00}$ or $K_A=T_{00}(H(x-\epsilon))$. Therefore,  it is more properly to define the local operator $P_{x,A}=T_{0x}(H(x-\epsilon))$ and $P_{x,\bar A}=T_{0x}(H(-x+\epsilon))$. One could show that $[P_{x,A},P_{x,\bar A}]=0$.\\

Now we could decompose the Hermitian operator as $\eta=\eta_{A}\otimes \eta_{\bar A}$ with  $\eta_A:= e^{\pi K_A} e^{i P_{A,x}(x_1+x_2)}$ and $\eta_{\bar A}:= e^{\pi K_{\bar A}} e^{i P_{\bar A,x}(x_1+x_2)}$. We also show $\eta_A$ and $\eta_{\bar A}$ are Hermitian and invertible operators. According to the theorem in \cite{} we conclude that the eigenvalues of $\transA$ will be real or complex coming in conjugated pairs, which is consistent with the pseudo R\'enyi entropy would be a real number (\ref{table:1}). \\

Now let us consider the two operators are not on same time slice. If  $O(t',x')$ is located in  right Rindler wedge, we will have the following relation
\bea\label{modularmap}
 O(t',x')|0\rangle=e^{-\pi K} O(-t',-x')|0\rangle,
\eea
where $O(-t',-x')$ is the operator located in the left Rindler wedge, $e^{-\pi K}$ is the modular operaor. By further using translation operators we have
\bea
O(t,x)|0\rangle \langle 0|O(t',x')=O(t,x)|0\rangle \langle 0|O(t,x)e^{-i H(t+t')+i P_x (x+x') }e^{-\pi K}.
\eea
Let us define the operator $\eta'=e^{-i H(t+t')+i P_x (x+x') }e^{-\pi K}$. By using the commutators (\ref{Poincare}), one could show $\eta'$ is a non-negative operator. Futher, we could decompose the operator $\eta'$ into local operators in $A$ and $\bar A$ as $\eta'_A=e^{-i H_A(t+t')+i P_{x,A} (x+x') }e^{-\pi K_A}$ and $\eta'_{\bar A}=e^{-i H_{\bar A}(t+t')+i P_{x,\bar A} (x+x') }e^{-\pi K_{\bar A}}$. We would have $\eta'=\eta'_A\otimes \eta'_{\bar A}$, where both $\eta'_A$ and $\eta'_{\bar A}$ are Hermitian and invertible. Thus the eigenvalues of $\transA$ are expected to be real or complex coming in conjugated pairs. The pseudo R\'enyi entropy should be real, which is consistent with our calculations. 

However,  if the operator $O(t',x')$ is located in the Kasner universe, the relation (\ref{modularmap}) is no longer right. For real $\theta$ the operator $e^{-2\pi i \theta K}$ is the Lorentz boost, which is unitary and acts on the state $O(t,x)|0\rangle$ as
\bea
e^{- i \theta K}O(t,x)|0\rangle=O(t',x')|0\rangle,
\eea
where $t'=t\cosh (\theta )+x\sinh (\theta)$ and $x'=t'=t\sinh (\theta)+x\cosh (\theta )$. Now we  would like to analytically continue $\theta$ to a complex parameter. If $\theta=i \pi$, one could obtain $t'=-t$ and $x'=-x$, thus we get the relation (\ref{modularmap}). But the analytical continuation is applicable only if $x>|t|$, that is in the Rindler wedge\cite{Witten:2018zxz}. Therefore, if $O(t',x')$ is located in the Kasner universe, the transition matrix (\ref{transitionlocal}) is no longer $\eta_A\otimes \eta_{\bar A}$-pseudo Hermitian. We expect the spetra of the reduced transition matrix would have be complex in general. The pseudo R\'enyi entropy would be complex. 

\section{Conclusions and discussions}

In this paper we investigate the pseudo R\'enyi entropy in QFTs. The transition matrix (\ref{transitionlocal}) is constructed by acting local operators on the vacuum.  We mainly focus on three different cases, in which the locations of the operator $O(t,x)$ are different, while $O(t',x')$ is fixed at the right Rindler wedge. 

In 2-dimensional CFTs we calculate the pseudo R\'enyi entropy for  some examples, including operators $\partial\phi \bar\partial \bar\phi$, $\mathcal{V}_\alpha$ in free scalar theory, $\phi_{(2,1)}$ in Minimal Models. It is found the pseudo R\'enyi entropy would be real for the cases \Rmnum{1} and \Rmnum{2}, that is the operator $O(t,x)$ is located at the left Rindler wedge. While the pseudo R\'enyi entropy is generally complex for case \Rmnum{3}, that is $O(t,x)$ is located at the Kanser universe. The results are summarized in the Table.\ref{table:1}.

Another interesting results in this paper is the universal divergent term of pseudo R\'enyi near the lightcone, i.e., $O(t,x)$ is located near the Rindler horizon $t\pm x=0$. The divergent behavior is observed in the paper \cite{Guo:2022sfl}, where the authors studied the time evolution of pseudo R\'enyi entropy. The situation is very similar, so our results are also used to understand the divergent behavior of the time evolution of pseudo R\'enyi entropy. It is found the second R\'enyi entropy shows the universal divergent term $2h\log|u|$, where $h$ is the conformal dimension of the operator $O$. The results are independent with the details of the theory, such as fusion rule of operators. Our results are only in 2-dimensional CFTs. The divergent behavior is closely related to the OPE of operators near the lightcone. By using the OPE of lightcone operators \cite{Brandt:1970kg} one may obtain more universal conclusions for general theory.

Finally, we use pseudo-Hermitian condition to explain the real-valued pseudo R\'enyi entropy in  case \Rmnum{1} and \Rmnum{2}. In the paper \cite{Guo:2022jzs} it is shown the real-valued condition can be associated with the pseudo-Hermiticity. Some examples are already discussed in \cite{Guo:2022jzs}. We further study some examples and extend the results in \cite{Guo:2022jzs}. For the case \Rmnum{1} and \Rmnum{2} we find the operator $\eta$, thus prove the pseudo-Hermitian condition for the reduced density matrix $\transA$. It should be noted  that the $\eta$-pseudo-Hermitian condition actually ensures the eigenvalues of the operator $\transA$ are real or complex coming in conjugated pairs. Thus the logarithmic term of the pseudo R\'enyi entropy are expected to be real for any $n$.  For the examples we actually only calculate pseudo R\'enyi entropy for $n=2,3$. Therefore, we could predict that the logarithmic term of $n$-th pseudo R\'enyi should be real for all the examples with the transition matrix (\ref{transitionlocal}) in case \Rmnum{1}  and \Rmnum{2}. One could check this prediction in more examples.

The pseudo R\'enyi entropy includes more information of the theory. It can be a useful probe to detect the  correlation functions, symmetry of the underlying theory. There are many interesing directions that are worth to explore in the near future. It is still unclear why the logarithmic  term of the pseudo R\'enyi entropy becomes complex for case \Rmnum{3}. It should be associated with the causality, which gives non-trival constraints on the correlators\cite{Hartman:2015lfa}. The divergent term of the pseudo R\'enyi entropy is also  mysterious. For the Rational CFTs one could understand the time evolution of R\'enyi entropy by using quasi-particles picture. This picture cannot be applied for the pseudo R\'enyi entropy. It is hard to image how the appearance of divergence by the quasi-particles picture, let alone the pseudo R\'enyi entropy may be complex. Recently, the authors in \cite{Guo:2023aio} find a sum rule for pseudo R\'enyi entropy. The pseudo R\'enyi entropy is associated with the R\'enyie entropy of the superposition state. By using this sum rule one may make more physical understanding of pseudo R\'enyi entropy in the quasi-particles picture. \\
~\\
~\\

{\bf Acknowledgements}
%%%%%%%%%%%%%%%%%%%%%%%%%%%%%
%%%%%%%%%%%%%%%%%%%%%%%%%%%%%%%%%%%%%
We would like to thank Xin Gao, Song He, Houwen Wu, Peng Wang, Haitang Yang, Long Zhao, Yu-Xuan Zhang and Zi-Xuan Zhao for valuable discussions related to this work.
WZG is supposed by the National Natural Science Foundation of China under Grant No.12005070 and the Fundamental Research Funds for the Central Universities under Grants NO.2020kfyXJJS041.

\appendix
\section{Cross ratio }\label{crossratiosection}
The variation of the second pseudo R\'enyi entropy $\Delta S^{(2)}$ is related to the cross ratio $\eta$ and $\bar \eta$. By analytical continuation  we have $w_1=x+t-i \epsilon$, $\bar w_1=x-t+ i \epsilon$  and $w_2=x'+t'-i \epsilon$, $\bar w_2=x'-t'+ i \epsilon$, where $\epsilon$ is the UV cut-off. In the final result we will take $\epsilon \to 0$. The cross ration is given by
\bea
&&\eta=\frac{1}{2}-\frac{t+t'+x+x'}{4 \sqrt{t+x-i \epsilon } \sqrt{t'+x'+i \epsilon }},\nn \\
&&\bar \eta=\frac{1}{2}-\frac{-t-t'+x+x'}{4 \sqrt{-t+x+i \epsilon } \sqrt{-t'+x'-i \epsilon }}.
\eea

For case I we have
\bea\label{crcaseI}
\eta=\frac{1}{2}-i\frac{x+x'}{4\sqrt{-x} \sqrt{x'}},\\
\bar \eta =\frac{1}{2}+ i\frac{x+x'}{4\sqrt{-x} \sqrt{x'}}. 
\eea
For case II we have
\bea\label{crcaseII}
&&\eta=\frac{1}{2}-i\frac{t+x+x'}{4\sqrt{-x-t} \sqrt{x'}},\\
&&\bar \eta=\frac{1}{2}+ i \frac{-t+x+x'}{4 \sqrt{t-x } \sqrt{x'}}.
\eea
For case III we have
\bea\label{crcaseIII}
&&\eta=\frac{1}{2}-\frac{t+x+x'}{4 \sqrt{t+x } \sqrt{x' }},\nn \\
&&\bar \eta=\frac{1}{2}+i\frac{-t+x+x'}{4 \sqrt{t-x } \sqrt{x' }}.
\eea
Note that near the light cone $t\sim -x$ the cross ratio $\eta$ would be divergent. If $t+x\to 0^-$ we have 
\bea\label{crlightconeneg}
&&\eta \to -\frac{i}{4\sqrt{x'}}(-t-x)^{-1/2},\nn \\
&&\bar \eta \to \frac{1}{2}+ i \frac{2x+x'}{4 \sqrt{-2x } \sqrt{x'}}.
\eea
If $t+x\to 0^+$ we have 
\bea\label{crlightconepos}
&&\eta\to -\frac{1}{4 \sqrt{x' }}(t+x)^{-1/2},\nn \\
&&\bar \eta \to \frac{1}{2}+i\frac{2x+x'}{4 \sqrt{-2x } \sqrt{x' }}
\eea

\section{More results of pseudo R\'enyi entropy for free boson theory}\label{appfreescalar}
Consider the operator $\p \phi \bar{\partial} \phi$. By
taking the cross ratio (\ref{crcaseII}) into (\ref{S2free1}) we obtain the result for case II
\bea\label{R2phicaseII}
&&\frac{tr (\mathcal{T}_A^{O|O'})^2}{tr(\rho_A^0)^2}\nn \\
&&=\left\{\frac{\left[-14 (t-x) x'+(t-x)^2+\left(x'\right)^2\right]\left[14 (t+x) x'+(t+x)^2+\left(x'\right)^2\right]}{256 \left(t^2-x^2\right) \left(x'\right)^2}\right\}^2,\nn \\
\eea
which is always positive. Similarly, taking (\ref{crcaseII}) into (\ref{S2free1}) we obtain the result for case III. The result is same as (\ref{R2phicaseII}). \\
For $n$-th pseudo R\'enyi entropy we can also directly evaluate the results by using Wick theorem. We obtain the results of $n=3$ and $n=4$ for case II as follows.

\bea\label{R3phicaseII}
&&\frac{tr (\mathcal{T}_A^{O|O'})^3}{tr(\rho_A^0)^3}\nn \\
&&=\left\{\frac{\left[-7(t-x)x'+(t-x)^2+(x')^2\right]\left[7(t+x)x'+(t+x)^2+(x')^2\right]}{81(t^2-x^2)(x')^2}\right\}^2,\nn \\
\eea
and
\begin{equation}\label{R4phicaseII}
	\begin{aligned}
		&\frac{tr (\mathcal{T}_A^{O|O'})^4}{tr(\rho_A^0)^4}\\
        &=\left[4096(t^2-x^2)(x')^2\right]^{-4}\left\{\left[(t-x)^2+62(x-t)x'+(x')^2\right]\right.\\
		&\left. \left[(t+x)^2+62(x+t)x'+(x')^2\right]\left[9(t-x)^2+46(x-t)x'+9(x')^2\right]\right.\\
        &\left.\left[9(t+x)^2+46(x+t)x'+9(x')^2\right]\right\}^2,
	\end{aligned}
\end{equation}
We can easily read the divergent behavior for the above two expressions near the lightcone.  For $n=3$ we find $tr (\mathcal{T}_A^{O|O'})^3/tr(\rho_A^0)^3\sim |u|^{-2}$, equally $\Delta S^{(3)}\sim \log |u|$. For $n=4$ we find $tr (\mathcal{T}_A^{O|O'})^4/tr(\rho_A^0)^4\sim |u|^{-4}$, equally $\Delta S^{(4)}\sim \frac{4}{3}\log |u|$. The results are consistent with the arguments in section.\ref{lightconeexamples}.

%We can generalize the above result for $\partial \phi \bar \partial \bar \phi$ to arbitrary $n$. 
%If $n$ is odd number, we can use 
\begin{figure}[H]
\centering
\subfloat[case \Rmnum{1}]{
		\includegraphics[scale=0.35]{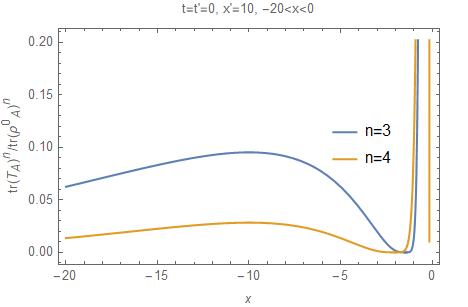}}
\subfloat[case \Rmnum{2}]{
		\includegraphics[scale=0.35]{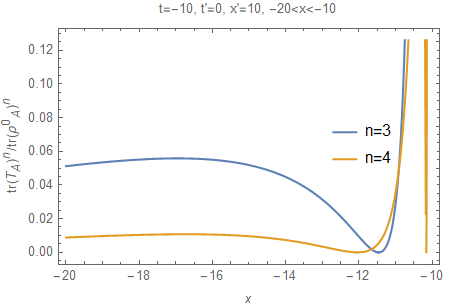}}
\\
\subfloat[case \Rmnum{2}]{
		\includegraphics[scale=0.38]{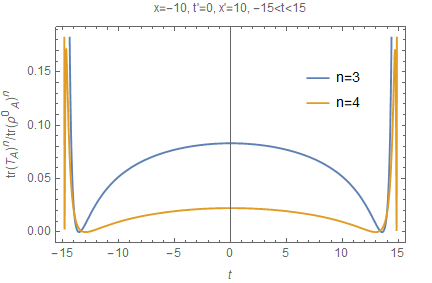}}
\subfloat[case \Rmnum{3}]{
		\includegraphics[scale=0.38]{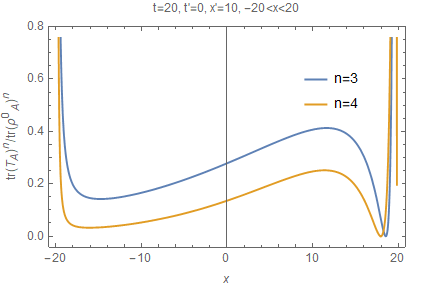}}
\caption{The plots of the logarithmic parts of $\Delta S^{(3)}$ and $\Delta S^{(4)}$ for the operator $\partial\phi\bar{\partial}\phi$. For all the plots $(t',x')$ is fixed to be $(0,10)$. (a) is for Case I, $t=0$, $x\in (-20,0)$. (b) is for Case II, $t=-10$, $x\in(-20,-10)$. (c) is also for Case II, $x=-15$, $t\in(-15,15)$. (d) is for Case III, $t=20$, $x\in(-20,20)$.}
\label{fig:doubleparphi,n=3,4}
\end{figure}

In the main text we show the second pseudo R\'enyi entropy  results for the operator $\mathcal{V}_\alpha$ with some fixed parameters $\alpha$.  Here we show the plot of the second pseudo R\'enyi entropy as a function of $\alpha$.
\begin{figure}[H]
\centering
\subfloat[Case \Rmnum{1}]{
		\includegraphics[scale=0.25]{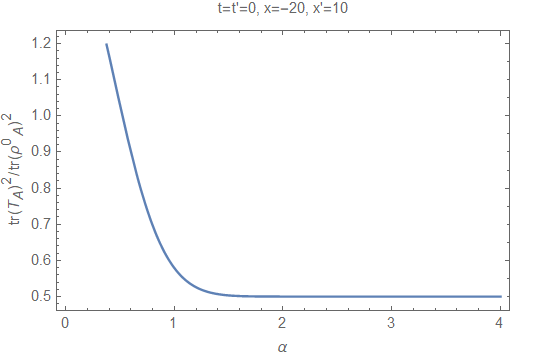}}
\subfloat[Case \Rmnum{2}]{
		\includegraphics[scale=0.25]{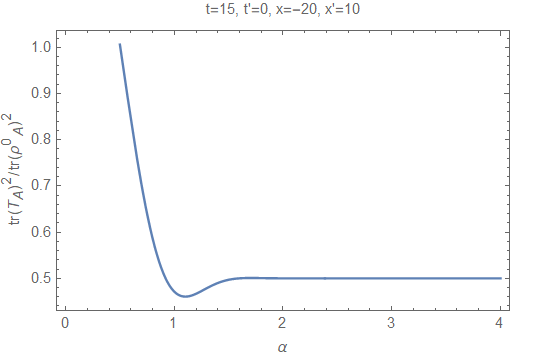}}
\\
\subfloat[Case \Rmnum{3}]{
		\includegraphics[scale=0.25]{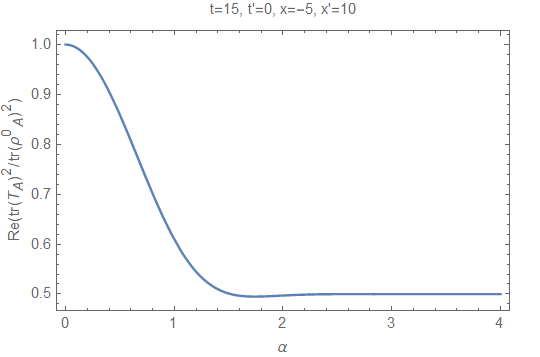}}
\subfloat[Case \Rmnum{3}]{
		\includegraphics[scale=0.25]{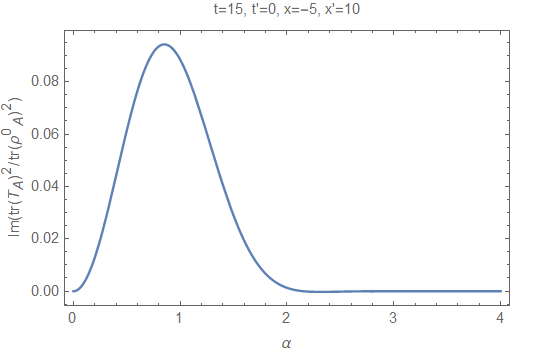}}
\caption{The plots of the logarithmic part of $\Delta S^{(2)}$ with respect to the parameter $\alpha$ in three cases. We fix $x'=10$. (a) is for Case \Rmnum{1}, $x=-20$, $\alpha\in (0,4)$, the results are real. (b) is for Case \Rmnum{2}, the result is also real. (c) and (d) are plots for the  real and imaginary parts for Case \Rmnum{3}. We take $t=15$ and $x=-5$.}
\label{fig:vertex,n=2,alpha}
\end{figure}

\par For the operator $\mathcal{V}_\alpha$ in $n\geq 3$, we can still calculate $\Delta S^{(n)}$ by using the formula (\ref{vertex,corfunction}), we show the expression for $n=3, \alpha=\frac{1}{2}$ in case II.

\begin{equation}
\frac{tr (\mathcal{T}_A^{O|O'})^3}{tr(\rho_A^0)^3}=\frac{\left [(-t-x)^{\frac{1}{3}}(t-x)^{\frac{1}{3}}+(x')^{\frac{2}{3}}\right ]^2}{4(-t-x)^{\frac{1}{3}}(t-x)^{\frac{1}{3}}(x')^{\frac{2}{3}}},
\end{equation}

for an arbitrary $\alpha$, we obtain $\frac{tr (\mathcal{T}_A^{O|O'})^3}{tr(\rho_A^0)^3}\sim{\lvert u\rvert}^{-\frac{4\alpha^2}{3}}$.

\begin{figure}[H]
\centering
\subfloat[case \Rmnum{1}]{
		\includegraphics[scale=0.35]{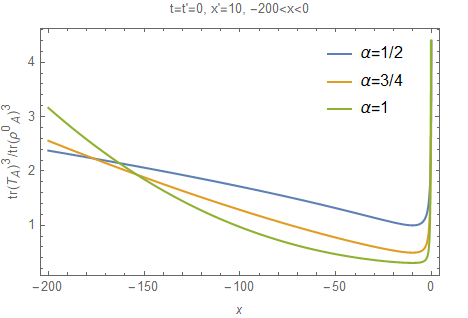}}
\subfloat[case \Rmnum{2}]{
		\includegraphics[scale=0.35]{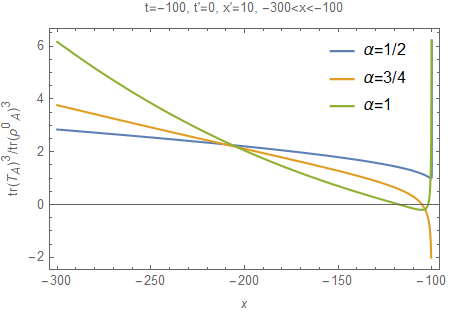}}
\\
\subfloat[case \Rmnum{3}]{
		\includegraphics[scale=0.35]{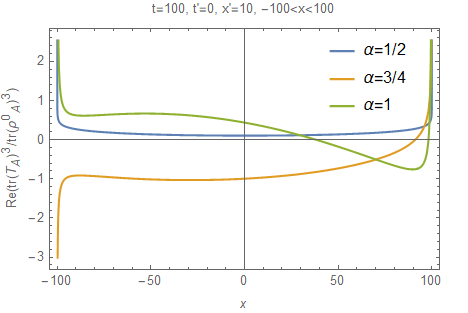}}
\subfloat[case \Rmnum{3}]{
		\includegraphics[scale=0.35]{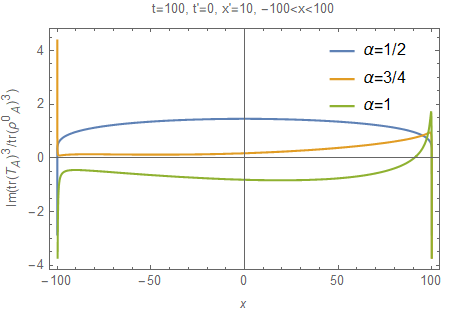}}
\caption{The plots of the logarithmic part of $\Delta S^{(3)}$ for $\mathcal{V}_\alpha$ in three cases. For all the plots $(t',x')$ is fixed to be $(0,10)$. (a) is for Case I, $t=0$, $x\in (-300,0)$. (b) is for Case II, $t=-100$, $x\in(-300,-100)$. (c) and (d) is also for Case III, $t=100$, $x\in(-100,100)$.}
\label{fig:vertex,n=3}
\end{figure}

\begin{figure}[H]
\centering
\subfloat[case \Rmnum{1}]{
		\includegraphics[scale=0.35]{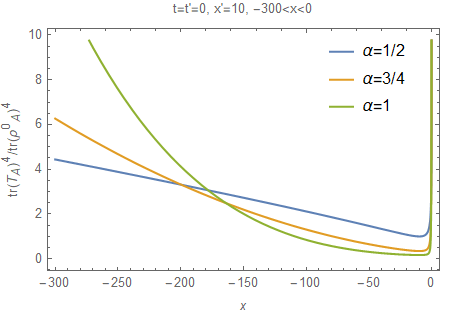}}
\subfloat[case \Rmnum{2}]{
		\includegraphics[scale=0.35]{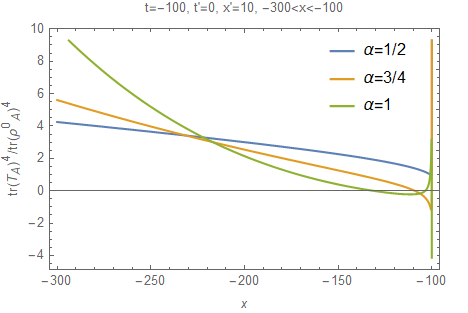}}
\\
\subfloat[case \Rmnum{3}]{
		\includegraphics[scale=0.35]{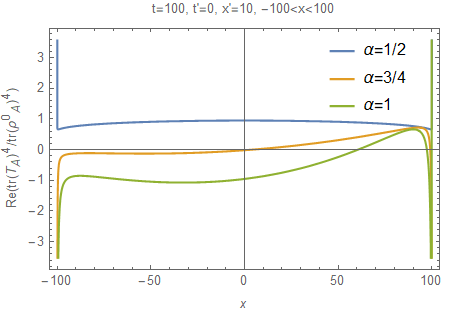}}
\subfloat[case \Rmnum{3}]{
		\includegraphics[scale=0.35]{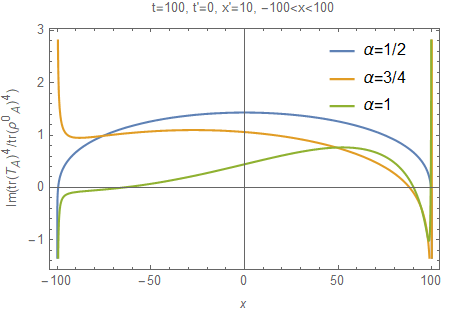}}
\caption{The plots of the logarithmic part of $\Delta S^{(4)}$ for $\mathcal{V}_\alpha$ in three cases.  For all the plots $(t',x')$ is fixed to be $(0,10)$. (a) is for Case I, $t=0$, $x\in (-300,0)$. (b) is for Case II, $t=-100$, $x\in(-300,-100)$. (c) and (d) is also for Case III, $t=100$, $x\in(-100,100)$.}
\label{fig:vertex,n=3}
\end{figure}

\section{Commutator of local operators for 2D CFTs}
Define the smearing operators 
\bea
T_{uu}(f):=\int T_{uu}(u)f(u)du,\quad T_{vv}(g):=\int T_{vv}(v)g(v)dv.
\eea
By using the commutator (\ref{stresscom}) we have
\bea
&&[T_{uu}(f),T_{uu}(f')]=i\int du T_{uu}\left(f(u)\p_u f'(u)-f'(u)\p_u f(u) \right)\nn \\
&&\phantom{[T_{uu}(f),T_{uu}(f')]=}-\frac{i c}{24\pi} \int du f(u)\p_u^3 f'(u)+i\int du \p_u (f(u)f'(u)T_{uu}).\nn
\eea
By using $T_{00}=T_{vv}+T_{uu}$ and $T_{0x}=T_{vv}-T_{uu}$, we can obtain the commutators
\bea\label{appendixcftstress}
&&[T_{00}(f),T_{00}(f')]=i \int dx T_{00}(0,x)\left(f(x)\p_x f'(x)-f'(x)\p_x f(x) \right)\nn \\
&&\phantom{[T_{00}(f),T_{00}(f')]=}+i\int dx \p_x (f(x)f'(x)T_{00}),\nn\\
&&[T_{00}(f),T_{0x}(f')]=i \int dx T_{00}(0,x)\left(f(x)\p_x f'(x)-f'(x)\p_x f(x) \right)\nn \\
&&\phantom{[T_{00}(f),T_{00}(f')]=}+i\int dx \p_x (f(x)f'(x)T_{00}),
\eea
where $T_{00}(f):=\int dx T_{00}(0,x)f(x)$, $T_{0x}(f):=\int dx T_{0x}(0,x)f(x)$. One could  also define the local operators 
\bea\label{localoperator}
&&H_A=T_{00}( H(x)),\quad  P_{A,x}=T_{0x}(H(x)),\quad K_A=T_{00}(x H(x)),\nn \\
&&H_{\bar A}=T_{00}( H(-x)),\quad  P_{\bar A,x}=T_{0x}(H(-x)),\quad K_{\bar A}=T_{00}(-x H(-x)).\nn \\
~
\eea
Using the commutators, we could obtain the relations
\bea
&&[K_A, H_A]=-i P_{A,x},\quad  [K_A, P_{A,x}]=-i H_A,\nn \\
&&[K_{\bar{A}}, H_{\bar A}]=i P_{\bar A,x},\quad  [K_{\bar A}, P_{\bar A,x}]=i H_{\bar A}.
\eea
One could also get
\bea
[K_A,P_{\bar A,x}]=0,\quad [K_{\bar A},P_{A,x}]=0,\quad [K_A,K_{\bar A}]=0.
\eea
By using the BCH formula, we have
\bea
&&e^{i\theta K_A}P_{A,x} e^{-i\theta K_A}= H_A \sinh \theta+P_{A,x} \cosh\theta\nn \\
&&e^{i\theta K_A}H_{A,x} e^{-i\theta K_A}=H_A\cosh\theta+P_{A,x}  \sinh \theta ,
\eea
which can be taken as a boost of the vector $(H_A,P_{A,x},0,0)$ in the $x$-direction. Taking $\theta=i\pi$, we have
\bea
e^{\pi K_A}P_{A,x} e^{-\pi K_A}=-P_{A,x},\quad e^{\pi K_A}H_{A} e^{-\pi K_A}=-H_{A}.
\eea

\section{Commutator of local operators for free scalar}\label{appfreecomu}
The Lagrangian density is given by
\bea
\mathcal{L}=-\frac{1}{2}\eta^{\mu\nu}\partial_\mu \phi \partial_\nu \phi-\frac{1}{2}m\phi^2.
\eea
The canonical stress energy tensor is 
\bea
T_{\mu\nu}=\partial_\mu\phi\partial_\nu \phi-\frac{1}{2}\eta_{\mu\nu}\partial^\sigma\phi \partial_\sigma \phi-\frac{1}{2}\eta_{\mu\nu}m\phi^2.
\eea
By using the commutator $[\phi(0,\vec{x}),\partial_0\phi(0,\vec{x}')]=i\delta(\vec{x}-\vec{x}')$, we have\footnote{One could check that $[H, T_{0i}(0,\vec{x}')]:=\int_Vd^3x[T_{00}(0,\vec{x}),T_{0i}(0,\vec{x}')]=-i \p_0 T_{0i}$, $[H, T_{00}(0,\vec{x}')]:=\int_Vd^3x[T_{00}(0,\vec{x}),T_{00}(0,\vec{x}')]=-i \p_0 T_{00}$ and   $[P_i, T_{00}(0,\vec{x}')]:=\int_Vd^3x[T_{0i}(0,\vec{x}),T_{00}(0,\vec{x}')]=-i \p_i T_{00}$. }
\bea\label{stressfree}
&&[T_{00}(0,\vec{x}),T_{00}(0,\vec{x}')]=-i \p_0\phi(0,\vec{x})\p^j \phi(0,\vec{x}')\p_j \delta(\vec{x}'-\vec{x})\nn \\
&&\phantom{[T_{00}(0,\vec{x}),T_{00}(0,\vec{x}')]=}+i \p_0\phi(0,\vec{x}')\p^j \phi(0,\vec{x})\p_j \delta(\vec{x}-\vec{x}').\nn\\
&&[T_{0i}(0,\vec{x}),T_{0i}(0,\vec{x}')]=-i \p_0 \phi(0,\vec{x}')\p_i \phi(0,\vec{x})\p_i \delta(\vec{x}'-\vec{x})\nn \\
&&\phantom{[T_{0i}(0,\vec{x}),T_{0i}(0,\vec{x}')]=}+i  \p_0 \phi(0,\vec{x})\p_i \phi(0,\vec{x}')\p_i \delta(\vec{x}-\vec{x}'),
\nn\\
&&[T_{00}(0,\vec{x}),T_{0i}(0,\vec{x}')]=-i \partial_0 \phi(0,\vec{x})\partial_0 \phi(0,\vec{x}')\partial_i \delta(\vec{x}'-\vec{x})\nn\\
&&+i\p^j \phi(0,\vec{x})\p_i \phi(0,\vec{x}') \p_j \delta(\vec{x}-\vec{x}')+im \phi(0,\vec{x})\p_i \phi(0,\vec{x}')\delta(\vec{x}-\vec{x}').
\eea
Using the commutators (\ref{stressfree}) one could evaluate the smearing operators,
\bea
T_{00}(f):= \int d^3x f(\vec{x})T_{00}(0,\vec{x}),\quad T_{0i}(g):= \int d^3x g(\vec{x})T_{0i}(0,\vec{x}),
\eea
where $f$ and $g$ are smooth functions. One could construct the local operators by choosing the support of these functions. We have
\bea\label{c1}
&&[T_{00}(f),T_{00}(f')]=-i\int d^3x f(\vec{x}) \int d^3x' f'(\vec{x}') \p_0\phi(0,\vec{x})\p^j \phi(0,\vec{x}')\p_j \delta(\vec{x}'-\vec{x})\nn \\
&&\phantom{[T_{00}(f),T_{00}(f')]=}+i\int d^3x f(\vec{x}) \int d^3x' f'(\vec{x}') \p_0\phi(0,\vec{x}')\p^j \phi(0,\vec{x})\p_j \delta(\vec{x}-\vec{x}')\nn\\
%&&\phantom{[T_{00}(f),T_{00}(f')]}=-i \int d^3x' f'(\vec{x}') \p^j \phi(0,\vec{x}')\p_j \int d^3x f(\vec{x})\p_0\phi(0,\vec{x}) \delta(\vec{x}'-\vec{x})\nn \\
%&&\phantom{[T_{00}(f),T_{00}(f')]=}+i\int d^3x f(\vec{x}) \p^j \phi(0,\vec{x})\p_j \int d^3x' f'(\vec{x}') \p_0\phi(0,\vec{x}') \delta(\vec{x}-\vec{x}').\nn\\
%&&\phantom{[T_{00}(f),T_{00}(f')]}=-i \int d^3x' f'(\vec{x}')\p_j f(\vec{x}') \p_0\phi(0,\vec{x}')\p^j \phi(0,\vec{x}') \nn \\
%&&\phantom{[T_{00}(f),T_{00}(f')]=}-i \int d^3x' f'(\vec{x}')f(\vec{x}')\p_j \p_0\phi(0,\vec{x}')\p^j \phi(0,\vec{x}') \nn \\
%&&\phantom{[T_{00}(f),T_{00}(f')]=}+i\int d^3x f(\vec{x})\p_j f'(\vec{x}) \p_0\phi(0,\vec{x}')\p^j \phi(0,\vec{x})\nn \\
%&&\phantom{[T_{00}(f),T_{00}(f')]=}+i\int d^3x f(\vec{x})f'(\vec{x})\p_j\p_0\phi(0,\vec{x})  \p^j \phi(0,\vec{x})\nn\\
&&\phantom{[T_{00}(f),T_{00}(f')]}=i\int d^3x [f(\vec{x})\p^j f'(\vec{x})-f'(\vec{x})\p^j f(\vec{x})] T_{0j}.
\eea
\bea\label{c3}
&&[T_{0i}(f),T_{0i}(f')]=-i \int d^3x' f'(\vec{x}')\p_j f(\vec{x}') \p_0\phi(0,\vec{x}')\p^j \phi(0,\vec{x}') \nn \\
&&\phantom{[T_{00}(f),T_{00}(f')]=}-i \int d^3x' f'(\vec{x}')f(\vec{x}')\p_j \p_0\phi(0,\vec{x}')\p^j \phi(0,\vec{x}') \nn \\
&&\phantom{[T_{00}(f),T_{00}(f')]=}+i\int d^3x f(\vec{x})\p_j f'(\vec{x}) \p_0\phi(0,\vec{x}')\p^j \phi(0,\vec{x})\nn \\
&&\phantom{[T_{00}(f),T_{00}(f')]=}+i\int d^3x f(\vec{x})f'(\vec{x})\p_j\p_0\phi(0,\vec{x})  \p^j \phi(0,\vec{x})\nn\\
&&\phantom{[T_{00}(f),T_{00}(f')]}=i\int d^3x [f(\vec{x})\p^j f'(\vec{x})-f'(\vec{x})\p^j f(\vec{x})] T_{0j}.
\eea
Similarly, we also have
\bea\label{c2}
&&[T_{00}(f),T_{0i}(f')]=-i\int d^3x f(\vec{x}) \int d^3x' f'(\vec{x}') \partial_0 \phi(0,\vec{x})\partial_0 \phi(0,\vec{x}')\partial_i \delta(\vec{x}'-\vec{x})\nn\\
&&\phantom{[T_{00}(f),T_{0i}(f')]=}+i\int d^3x f(\vec{x}) \int d^3x' f'(\vec{x}')\p^j \phi(0,\vec{x})\p_i \phi(0,\vec{x}') \p_j \delta(\vec{x}-\vec{x}')\nn \\
&&\phantom{[T_{00}(f),T_{0i}(f')]=}+im\int d^3x f(\vec{x}) \int d^3x' f'(\vec{x}') \phi(0,\vec{x})\p_i \phi(0,\vec{x}')\delta(\vec{x}-\vec{x}')\nn \\
&&\phantom{[T_{00}(f),T_{0i}(f')]}=i \int d^3x    f(\vec{x})f'(\vec{x})\p_i T_{00}+i \int d^3x   f(\vec{x})\p_i f'(\vec{x})  T_{00}\nn \\
&&\phantom{[T_{00}(f),T_{0i}(f')]=}+i \int d^3x    f(\vec{x})\p^j f'(\vec{x}) T_{ji}-i\int d^3x   \p_i[ f(\vec{x}) f'(\vec{x}) (\p_0 \phi)^2]\nn \\
&&\phantom{[T_{00}(f),T_{0i}(f')]}=i \int d^3x    f(\vec{x})\p^j f'(\vec{x}) T_{ji}-i \int d^3x \p_i f(\vec{x}) f'(\vec{x})  T_{00}\nn\\
&&\phantom{[T_{00}(f),T_{0i}(f')]=}- i\int d^3x \p_i[f(\vec{x})f'(\vec{x}) \mathcal{L}],
\eea
where the term in the last line is total derivative, which can be written as a boundary term using integral by parts.\\
 
Let us consider the commutators between the local operators and $\phi$, $\pi$.  With some calculations for operator $\phi$ located in the right Rindler wedge we have
\bea
e^{i \theta K}\phi(0,x,y,z)e^{-i \theta K}=\phi(t(\theta),x(\theta),y,z),
\eea
where $t(\theta)=t \cosh \theta+x \sinh \theta$ and $x(\theta)=x \cosh \theta+t \sinh \theta$. $K_A$ generate the boost in $x$-direction. If taking $\theta=i\pi$, we have $t(i\pi)=-t$ and $x(i\pi)=-x$, that is the operator is mapped into the left Rindler wedge by $e^{-\pi K}$. \\
$P_{x}$ generates the translation along $x$-direction, we have
\bea
e^{i P_{x} a}\phi(0,x,y,z)e^{-i P_{x} a}=\phi(0,x+a,y,z),
\eea
if $\phi$ is located in the region $A$.

\end{document}